\documentclass{revtex4}
\usepackage{graphicx}
\begin{document}
\title{Rogue waves statistics in the framework of one-dimensional Generalized Nonlinear Schrodinger Equation}

\author{D.S. Agafontsev$^{(a)}$, V.E. Zakharov$^{(a),(b),(c)}$}
\affiliation{\small \textit{ $^{(a)}$ P. P. Shirshov Institute of Oceanology, 36 Nakhimovsky prosp., Moscow 117218, Russia\\
$^{(b)}$ L. D. Landau Institute for Theoretical Physics, 2 Kosygin str., 119334 Moscow, Russia\\
$^{(c)}$ Department of Mathematics, University of Arizona, Tucson, AZ, 857201, USA}}

\begin{abstract}
We measure evolution of spectra, spatial correlation functions and probability density functions (PDFs) of waves appearance for a set of one-dimensional NLS-like equations of focusing type, namely for the classical integrable Nonlinear Schrodinger equation (1), nonintegrable NLS equation accounting for dumping (linear dissipation, two- and three-photon absorption) and pumping terms (2) and generalized NLS equation accounting for six-wave interactions, dumping and pumping terms (3). All additional terms beyond the classical NLS equation are small. As initial conditions we choose seeded by noise modulationally unstable solutions of the considered systems in the form of (a) condensate for systems (1)-(3) and (b) cnoidal wave for the classical NLS equation (1). We observe 'strange' results for the classical NLS equation (1) with condensate initial condition including peak at zeroth harmonic in averaged over ensemble spectra, non-decaying spacial correlation functions and a 'breathing' region on the PDFs for 
medium waves amplitudes where frequency of waves appearance oscillates with time, while the far-tails of the PDFs remain Rayleigh ones. Addition of small dumping and pumping terms in model (2) breaks integrability that results in absence of the peak at zeroth harmonic in spectra, spacial correlation functions decaying to zero level and strictly Rayleigh PDFs for waves amplitudes. For the classical NLS equation (1) with cnoidal wave initial condition PDFs turn out to be significantly different from Rayleigh ones with 'fat tails' in the region of large amplitudes where higher waves appear more frequently, while for generalized NLS equation with six-wave interactions, dumping and pumping terms (3) we demonstrate absence of non-Rayleigh addition to the PDFs for zeroth six-wave interactions coefficient and increase of non-Rayleigh addition with six-wave interactions term.
\end{abstract}

\maketitle


{\bf 1. Introduction.} \\

Since the first observation by Solli \textit{et al.} in 2007 \cite{Solli}, optical rogue waves - large wave events that appear randomly from initially smooth pulses and have statistics drastically different from that predicted by the linear theory - has drawn much scientific attention from both optical and hydrodynamic society. In case of optics rogue waves are huge waves that can damage optical systems and therefore their appearance must be controlled. For hydrodynamics optical rogue waves are interesting phenomenon that can be conveniently studied in laboratory conditions and that occurs in systems described by the similar equations and has similar statistics to hydrodynamic rogue waves \cite{Dudley3}. The current study of optical rogue waves went in two main directions: 1) harnessing and control of rogue waves emergence by shaping the initial pulse \cite{Solli2, Dudley1}, and 2) understanding the physical mechanisms underlying the phenomenon by establishing connection between different linear and 
nonlinear terms in the equations of motion and appearance of singularities in the probability of large waves occurrence, especially so-called 'fat tails' when higher waves appear by several orders of magnitude more frequently comparing to the linear theory \cite{Dudley2, Taki, Picozzi, Agafontsev, Hadzievski, Passot}. In the current publication we implement both approaches as we compare statistics for integrable Nonlinear Schrodinger (NLS) equation with that for nonintegrable one with small additional dumping and pumping terms and study the influence of different initial conditions and higher nonlinearity (six-wave interactions) on the frequency of rogue waves appearance.

Investigation of the influence of initial conditions is necessary because different types of initial conditions contain different fractions of solitons and linear waves and the statistics of solitons as stable essentially nonlinear objects may be very different from that for the linear waves. In this respect cnoidal waves are very convenient objects to study because increasing its imaginary half-period allows one to move from mostly solitonic states to initial conditions with significant fraction of linear waves. We perform comparison between integrable NLS equation and its nonintegrable conterpart with only slight modifications beyond the classical NLS equation because from one hand the classical integrable NLS equation conserves infinite number of integrals of motion and we might expect some unique behavior therefore, while from the other hand in the real physical systems some additional terms are always present. Concerning the effect of higher nonlinearity our research can also be considered in the 
broader context. The classical NLS equation generalizations of which are often studied in the framework of rogue waves phenomenon, describes many physical systems from Bose-Einstein condensate and plasma oscillations to propagation of waves in optics and hydrodynamics. We consider generalized NLS equation with six-wave interactions term that naturally appears as next-order term in perturbation theory expansion and investigate its influence on the map of extreme events of the system in the regime when dynamics of the system is close to that described by the classical NLS equation. In this sense our choice of parameters is more natural than in the recent publication \cite{Lushnikov} where the similar research was made for the quintic NLS equation.

\begin{figure}[h] \centering
\includegraphics[width=130pt]{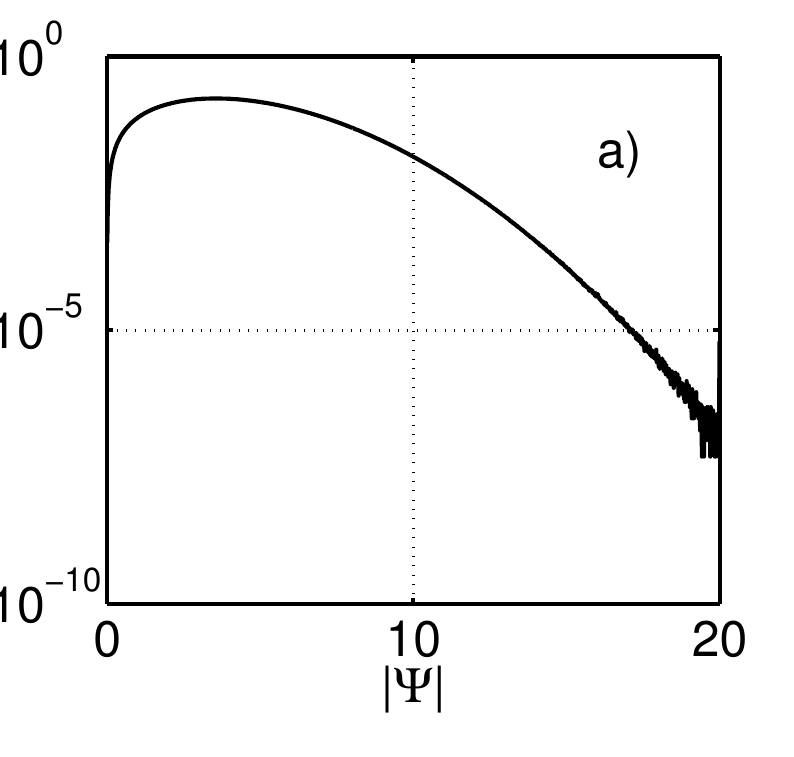}
\includegraphics[width=130pt]{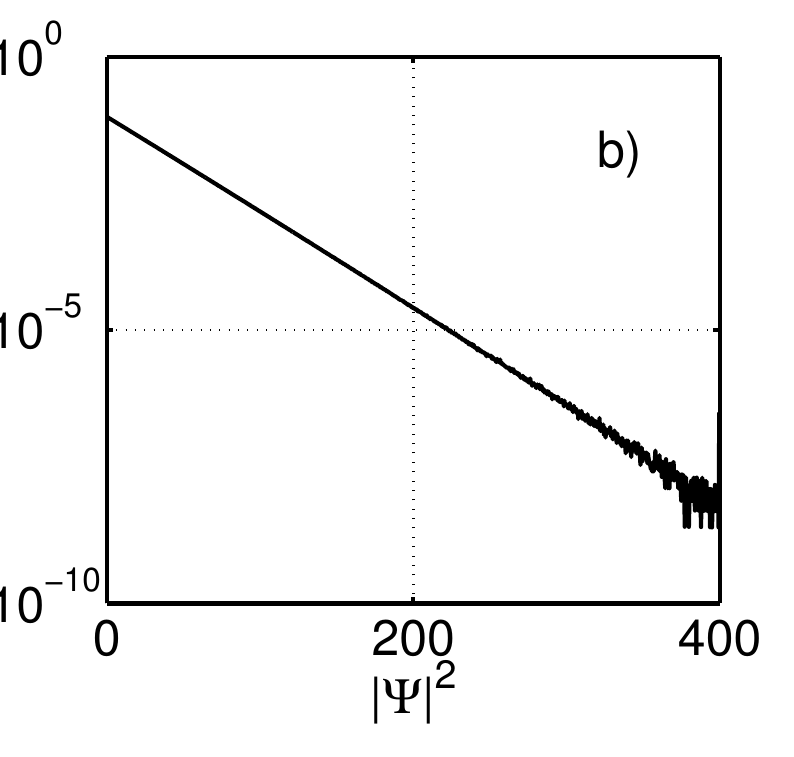}
\caption{\small {\it Normalized PDF for linear waves $\Psi(x) = 0.5\int_{-\infty}^{+\infty}\exp(-0.04 k^{2} + i\phi_{k})\exp(ikx)(dk/\sqrt{2\pi})$ in log-scale calculated on $10^{6}$ different realizations of random uncorrelated values $\phi_{k}$: (a) - $PDF(|\Psi|)$ depending on $|\Psi|$, (b) - $PDF(|\Psi|)/|\Psi|$ depending on $|\Psi|^{2}$.}}
\end{figure}

Let us suppose that the current state of a system consists of multitude of uncorrelated linear waves,
$$
\Psi = \sum_{k}a_{k}\, \exp(i(kx-\omega_{k}t+\phi_{k})).
$$
If $a_{k}$ and $\phi_{k}$ are random uncorrelated values and the number of linear waves is large enough, then probability to meet amplitude $|\Psi|$ (probability density function, PDF) obeys Rayleigh distribution (see example on FIG. 1a,b),
$$
PDF(|\Psi|) \sim |\Psi|\exp(-|\Psi|^{2}/2\sigma^{2}).
$$
In the current publication we search for deviations from Rayleigh distribution at large amplitudes, especially for signs of 'fat tails' when large waves occur much more frequently and therefore the map of extreme events is qualitatively different from that of a linear system. 

Because of the equality $\int_{0}^{+\infty} F(x)x\exp(-x^{2}/2\sigma^{2})\,dx = (1/2)\int_{0}^{+\infty} F(x)\exp(-x^{2}/2\sigma^{2})\,d\,x^{2}$ PDF for squared amplitudes $|\Psi|^{2}$, that is by definition the probability to meet a given squared amplitude $|\Psi|^{2}$, is exponential:
$$
PDF(|\Psi|^{2}) \sim \exp(-|\Psi|^{2}/2\sigma^{2}).
$$
Since it is obviously more convenient to examine exponential $\exp(-x), \, x=|\Psi|^{2},$ dependencies than Rayleigh $x\exp(-x^{2}), \, x=|\Psi|,$ ones, in the current publication we measure PDFs for squared amplitudes $|\Psi|^{2}$ instead of PDFs for amplitudes $|\Psi|$. As explained above, our results for squared amplitudes PDFs are easily translated to amplitudes PDFs: if a squared amplitudes PDF is exponential $PDF(|\Psi|^{2})\sim\exp(-|\Psi|^{2}/2\sigma^{2})$ then the corresponding amplitudes PDF is Rayleigh-distributed $PDF(|\Psi|) \sim |\Psi|\exp(-|\Psi|^{2}/2\sigma^{2})$ and vice versa, and for simplicity below we will call both such PDFs as Rayleigh-distributed illustrating them with graphs of squared amplitudes PDFs. Here we work with PDFs for entire field $|\Psi|^{2}$ only because it turned out that PDFs for local maximums did not give us any extra information.

In the current study we perform numerical simulations of the evolution of wave field $\Psi$ in the framework of three different nonlinear equations for ensembles of at least 10 000 initial distributions $\Psi|_{t=0}$ for each of the nonlinear systems. Inside each ensemble initial distributions differ only by realizations of stochastic noise with a fixed statistical properties. Based on the ensembles, we measure spectra $I_{k}=\langle|\Psi_{k}|^{2}\rangle$ where $\langle..\rangle$ stands for averaging over ensemble, spatial correlation functions $g(x)=\langle\Psi(y,t)\Psi^{*}(y+x,t)\rangle$ and PDFs for squared amplitudes $|\Psi|^{2}$ and examine how these functions depend on time, type of the initial condition and different nonlinear terms included in the equations of motion. 

The paper is organized as follows. In Section 2a we compare integrable classical NLS equation and nonintegrable one with small dumping and pumping terms for condensate initial condition, while Section 2b contains our results for the classical NLS equation for cnoidal wave initial conditions. Sections 3 is devoted to investigation of the generalized NLS equation with six-wave interactions, dumping and pumping terms. Section 4 contains conclusions and acknowledgements, while a brief overview of the numerical methods we used is given in the appendix.\\


{\bf 2a. Comparison of the classical integrable NLS equation and nonintegrable NLS equation with small dumping and pumping terms for condensate initial condition.} \\

In the current publication we study statistical properties of the solutions for the classical integrable Nonlinear Schrodinger equation of the focusing type,\\
\begin{equation}\label{Eq01}
i\Psi_t +\Psi_{xx} - \Psi +|\Psi|^2 \Psi = 0,
\end{equation}
and its generalizations accounting for higher nonlinearity, pumping and dumping terms. Here $t$ is time and $x$ is spatial coordinate. We consider the focusing four-wave interactions case only that corresponds to anomalous group velocity dispersion (GVD) regime because optical rogue waves are huge pulses lying entirely in the anomalous GVD \cite{Solli, Dudley3, Solli2, Dudley1, Dudley2, Taki, Picozzi, Agafontsev}.

Eq. (\ref{Eq01}) is the Hamiltonian one,
$$
i\Psi_t = \frac{\delta H}{\delta \Psi^{*}},
$$
with Hamiltonian 
$$
H = H_{d} + H_{4},
$$
where
$$
H_{d}=\int |\Psi_{x}|^2\,dx, \quad H_{4} = -\int\frac{|\Psi|^4}{2}\,dx.
$$
The classical Nonlinear Schrodinger equation (\ref{Eq01}) is integrable in the framework of Inverse Scattering Method \cite{Zakharov0} and has infinite number of integrals of motion, first three of them are wave action $N=\int|\Psi|^{2}\,dx$, momentum $P=(i/2)\int(\Psi_{x}^{*}\Psi-\Psi_{x}\Psi^{*})\,dx$, and Hamiltonian. Because of integrability we might expect some unique behavior for this equation and in order to check it we compare statistical results for Eq. (\ref{Eq01}) with that for its nonintegrable counterpart accounting for small dumping and pumping terms,
\begin{eqnarray}\label{Eq01_1}
& i\Psi_t +(1-id_{l})\Psi_{xx}-\Psi+(1+id_{2p})|\Psi|^2 \Psi +id_{3p} |\Psi|^4 \Psi = ip\Psi,\\ 
& d_{l},d_{2p},d_{3p},p>0, \quad d_{l},d_{2p},d_{3p},p \ll 1 \nonumber.
\end{eqnarray}
Eq. (\ref{Eq01_1}) takes into account linear dissipation (proportional to $d_{l}$) that may originate from optical filtering \cite{Renninger} and also nonlinear dissipation terms in the form of two- (proportional to $d_{2p}$) and three-photon (proportional to $d_{3p}$) absorption. In order to balance the system we also introduce deterministic pumping term $ip\Psi$. We checked other homogeneous in x-space forcing terms including chaotic forcing and found no significant difference in our results. Coefficients before all the additional terms in Eq. (\ref{Eq01_1}) beyond the classical NLS equation are chosen to be small so that the dynamics of Eq. (\ref{Eq01_1}) is close to that of the classical NLS equation. We checked several sets of coefficients $d_{l}$, $d_{2p}$, $d_{3p}$ and $p$ and found that our results for averaged spectra, spacial correlation functions and the PDFs do not significantly depend on them for a wide range of values. Therefore we fixed such dumping and pumping coefficients $d_{l}=0.0324$, $d_{
2p}=0$, $d_{3p}=0.0001$, $p=0.02$ that lead to the same mean wave action, total, kinetic and potential energy (that are all functions of time for Eq. (\ref{Eq01_1})) as for the integrable case (\ref{Eq01}).

We solve Eq. (\ref{Eq01})-(\ref{Eq01_1}) numerically starting from the initial data $\Psi|_{t=0}=1+\epsilon(x)$, where an exact modulationally unstable solution of Eq. (\ref{Eq01}) in the form of condensate $\Psi|_{t=0}^{(0)}=1$ is seeded by stochastic noise $\epsilon(x)$ with fixed statistical properties. For all of the nonlinear systems examined in this publication we used noise in the form of Gaussian-distributed in k-space function $\epsilon(x)=A_{0} \int exp(-k^{2}/\theta^{2}+i\xi_{k})\,exp(ikx)\,(dk/\sqrt{2\pi})$ with arbitrary phases $\xi_{k}$ and with relatively large dispersion $\theta>>k_{0}$ exceeding wavenumber $k_{0}=1$ corresponding to maximum increment of modulation instability. Amplitude of noise was chosen sufficiently small $A_{0}\sim 10^{-5}-10^{-3}$ in order to ensure that the deviations in initial wave action, energy and momentum inside the ensembles were small. We did not find difference in our results using other homogeneous in x-space statistical distributions of noise. 

\begin{figure}[h] \centering
\includegraphics[width=130pt]{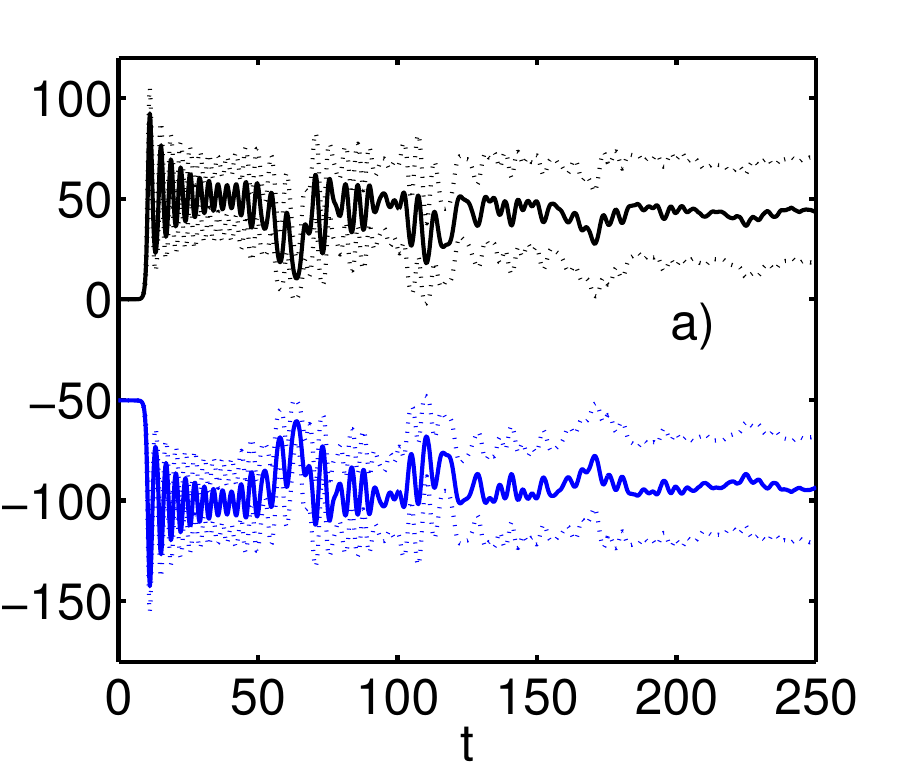}
\includegraphics[width=130pt]{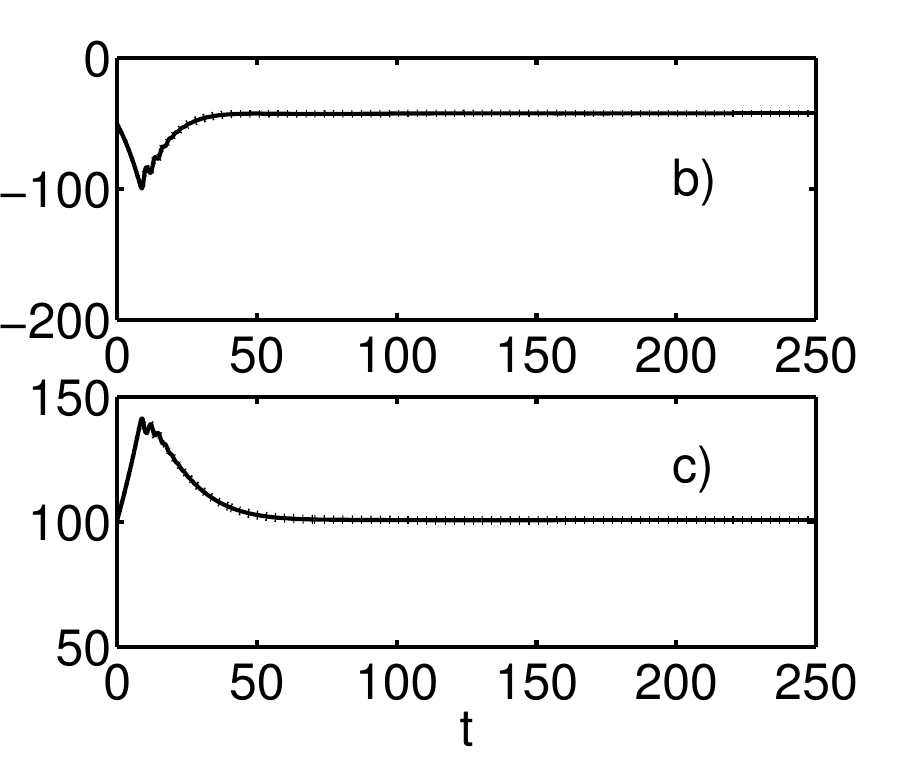}
\includegraphics[width=130pt]{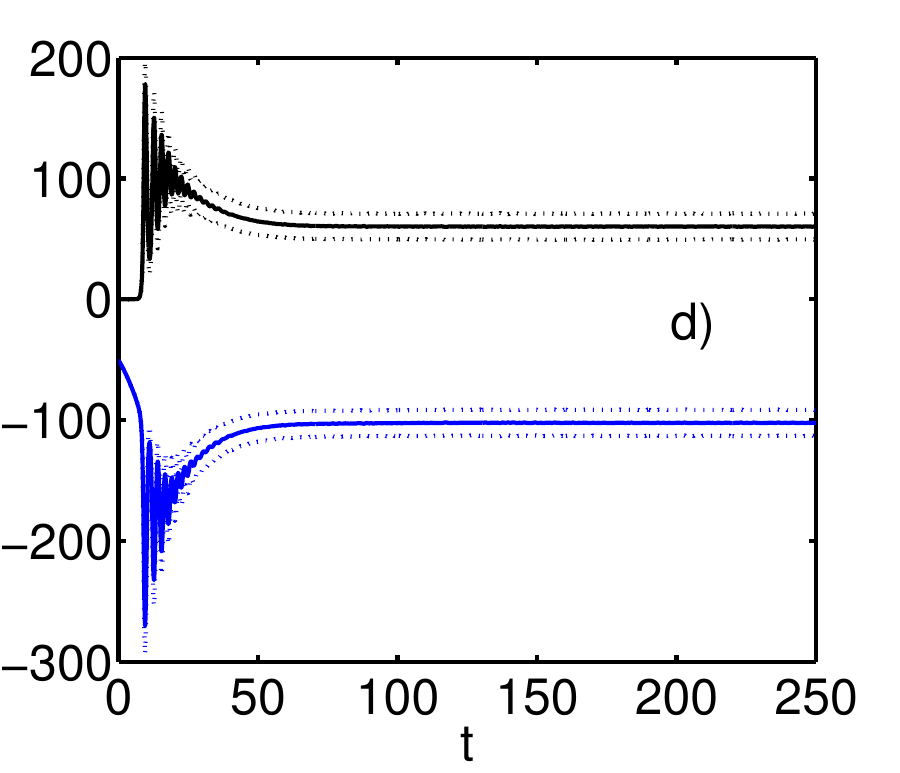}
\caption{\small {\it  (Color on-line) Evolution of averaged over ensemble kinetic $\langle H_{d}\rangle$ (black) and potential $\langle H_{4}\rangle$ (blue) energy for the classical NLS equation (\ref{Eq01}) for condensate initial condition (a); evolution of wave action $N$ (b), energy $H$ (c), kinetic $\langle H_{d}\rangle$ (black) and potential $\langle H_{4}\rangle$ (blue) energy (d) for the generalized NLS equation accounting for dumping and pumping terms (\ref{Eq01_1}), $d_{l}=0.0324$, $d_{2p}=0$, $d_{3p}=0.0001$, $p=0.02$ for condensate initial condition. Solid lines - mean over ensemble values, dashed lines - borders for the corresponding standard deviations.}}
\end{figure}

\begin{figure}[h] \centering
\includegraphics[width=130pt]{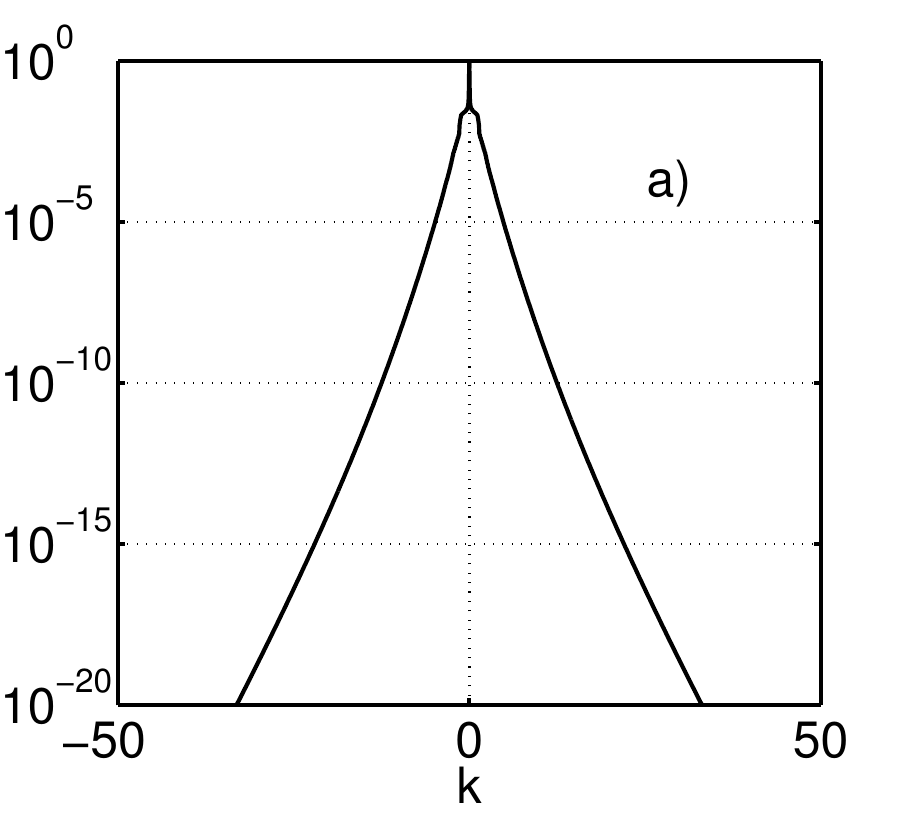}
\includegraphics[width=130pt]{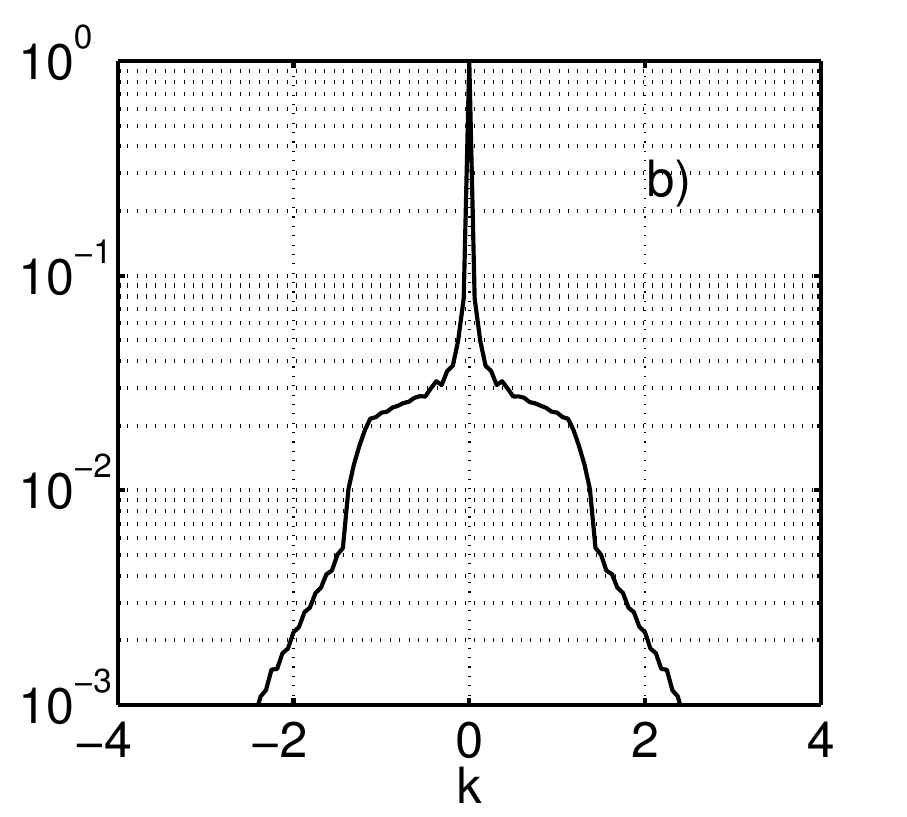}
\includegraphics[width=130pt]{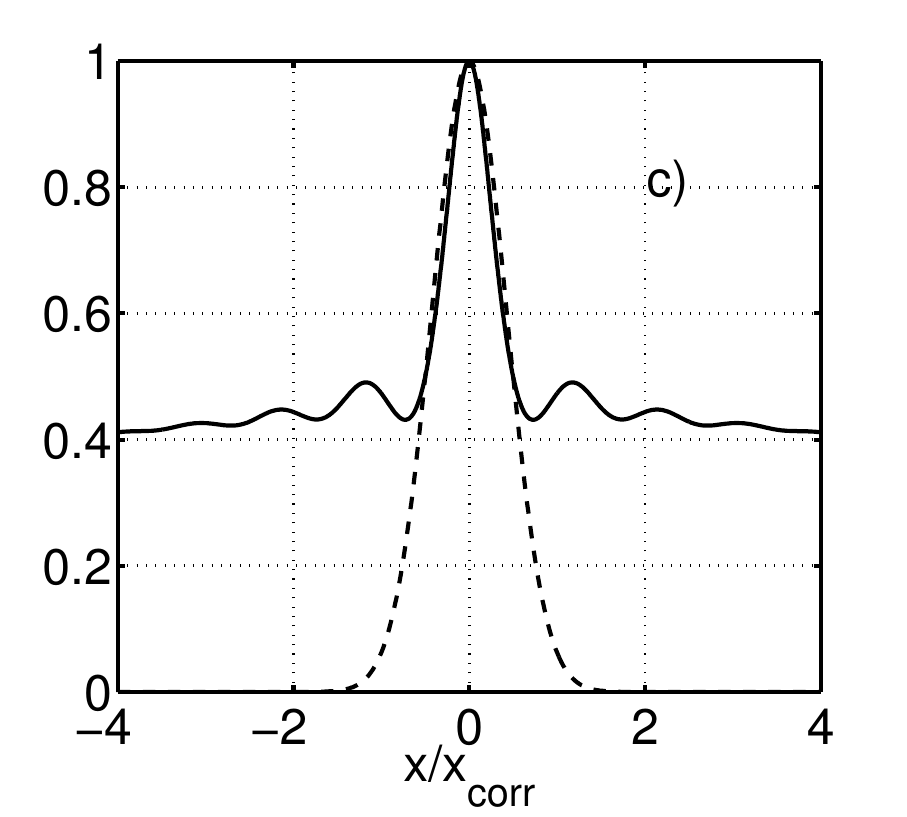}
\caption{\small {\it  Averaged over ensemble and time normalized spectra $I_{k}/\max I_{k}$ in full scale (a) and enlarged at center region (b) as well as spatial correlation functions $g(x/x_{corr})/g(0)$ (c) for condensate initial condition for the classical NLS equation (\ref{Eq01}) at $t\in [10, 250]$. For graph (c) dashed line is Gaussian distribution.}}
\end{figure}

\begin{figure}[h] \centering
\includegraphics[width=320pt]{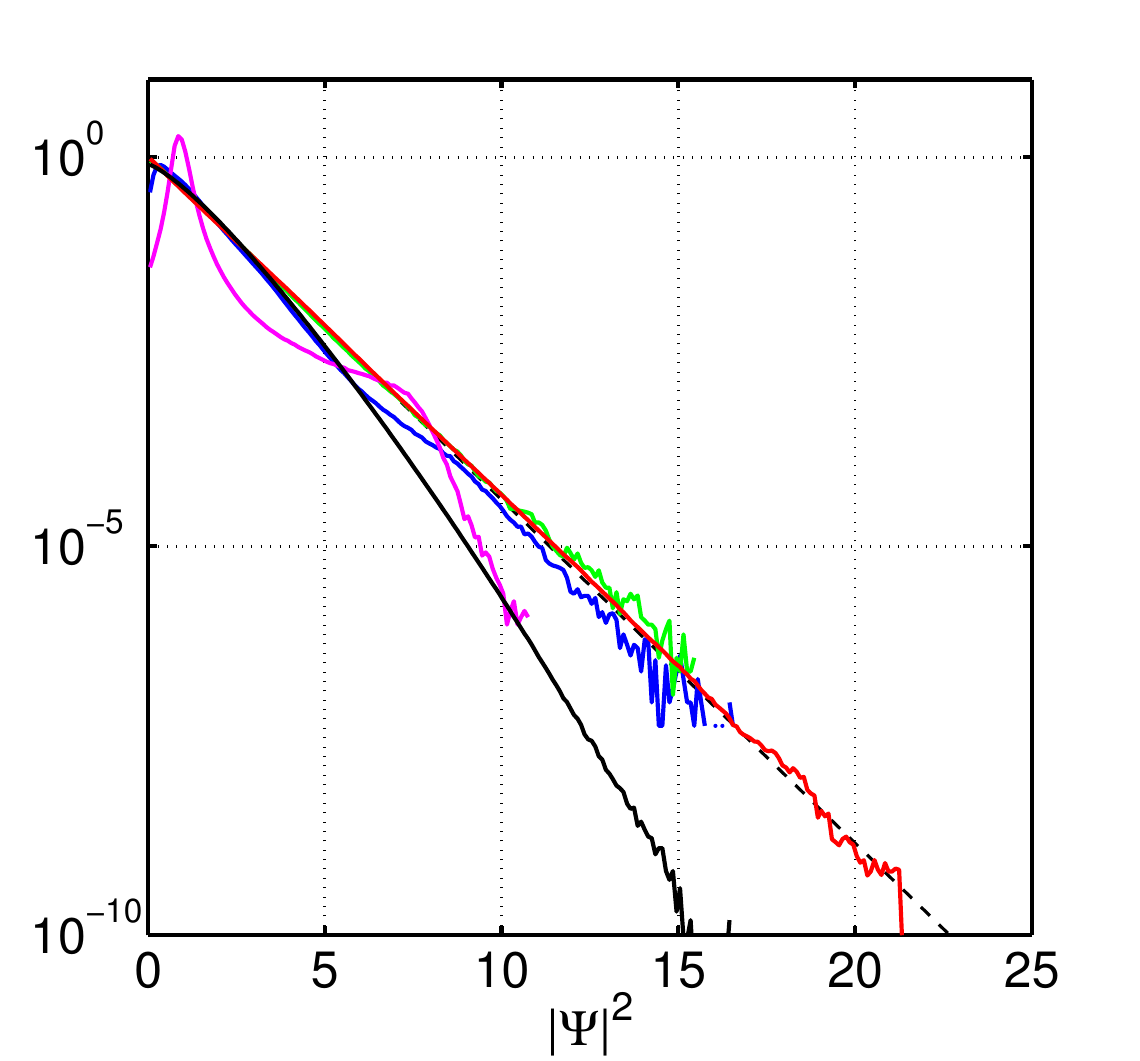}
\caption{\small {\it  (Color on-line) Averaged over ensemble normalized squared amplitudes PDFs in semi-log scale for condensate initial condition for the classical NLS equation (\ref{Eq01}). Dashed line is exponential $\sim\exp(-|\Psi|^{2}/2\sigma^{2})$ dependency with $\sigma\approx 0.86$. Blue line corresponds to $t=14$, green - to $t=30$, purple - to $t=65$, red - to time-averaged at $t\in [10, 250]$ PDF, black - to time-averaged PDF of a linear system with the same spectra $I_{k}$ as for the classical NLS equation (\ref{Eq01}).}}
\end{figure}

\begin{figure}[h] \centering
\includegraphics[width=130pt]{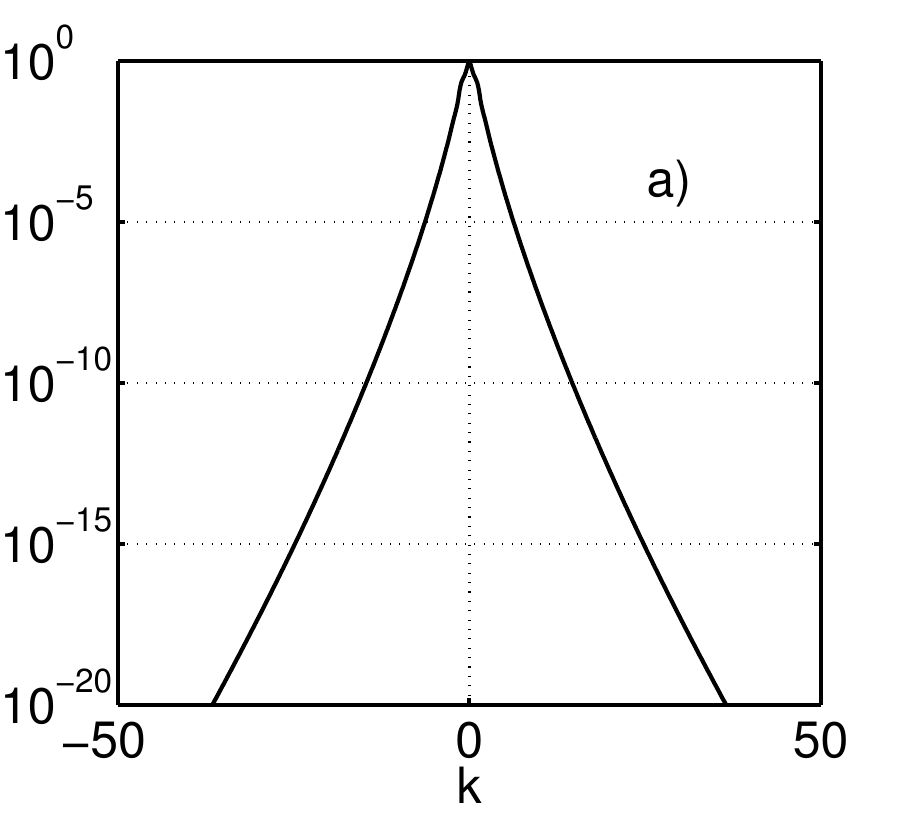}
\includegraphics[width=130pt]{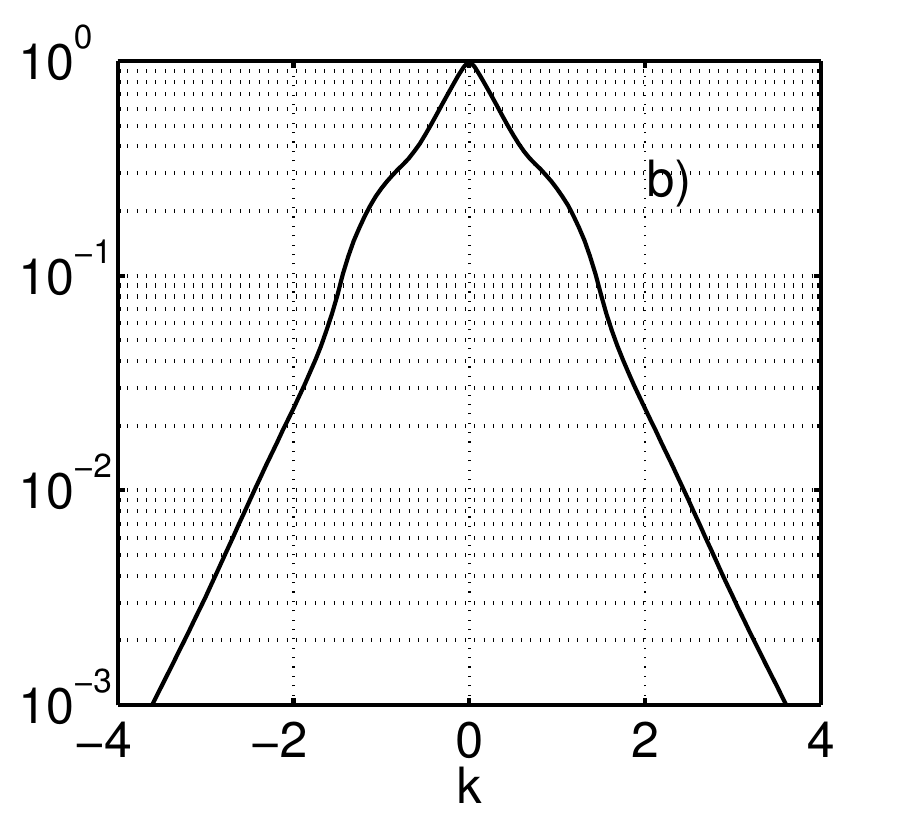}
\includegraphics[width=130pt]{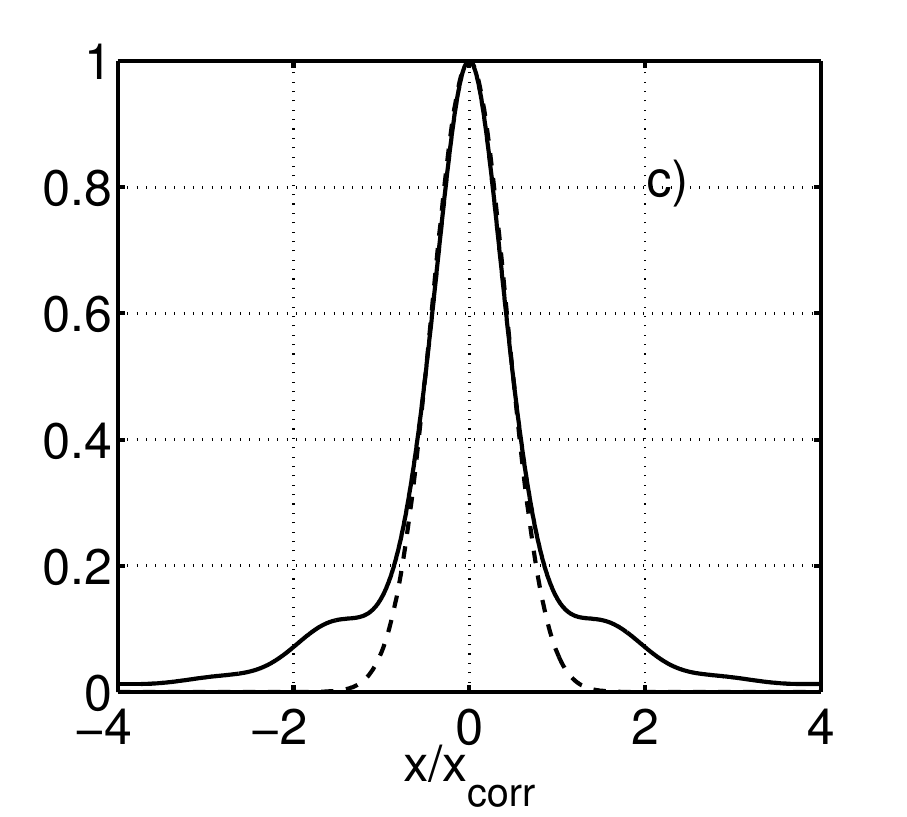}
\caption{\small {\it  Averaged over ensemble and time normalized spectra $I_{k}/\max I_{k}$ in full scale (a) and enlarged at center region (b) as well as spatial correlation functions $g(x/x_{corr})/g(0)$ (c) for condensate initial condition for generalized NLS equation accounting for dumping and pumping terms (\ref{Eq01_1}) at $t\in [50, 250]$, $d_{l}=0.0324$, $d_{2p}=0$, $d_{3p}=0.0001$, $p=0.02$. For graph (c) dashed line is Gaussian distribution.}}
\end{figure}

\begin{figure}[h] \centering
\includegraphics[width=320pt]{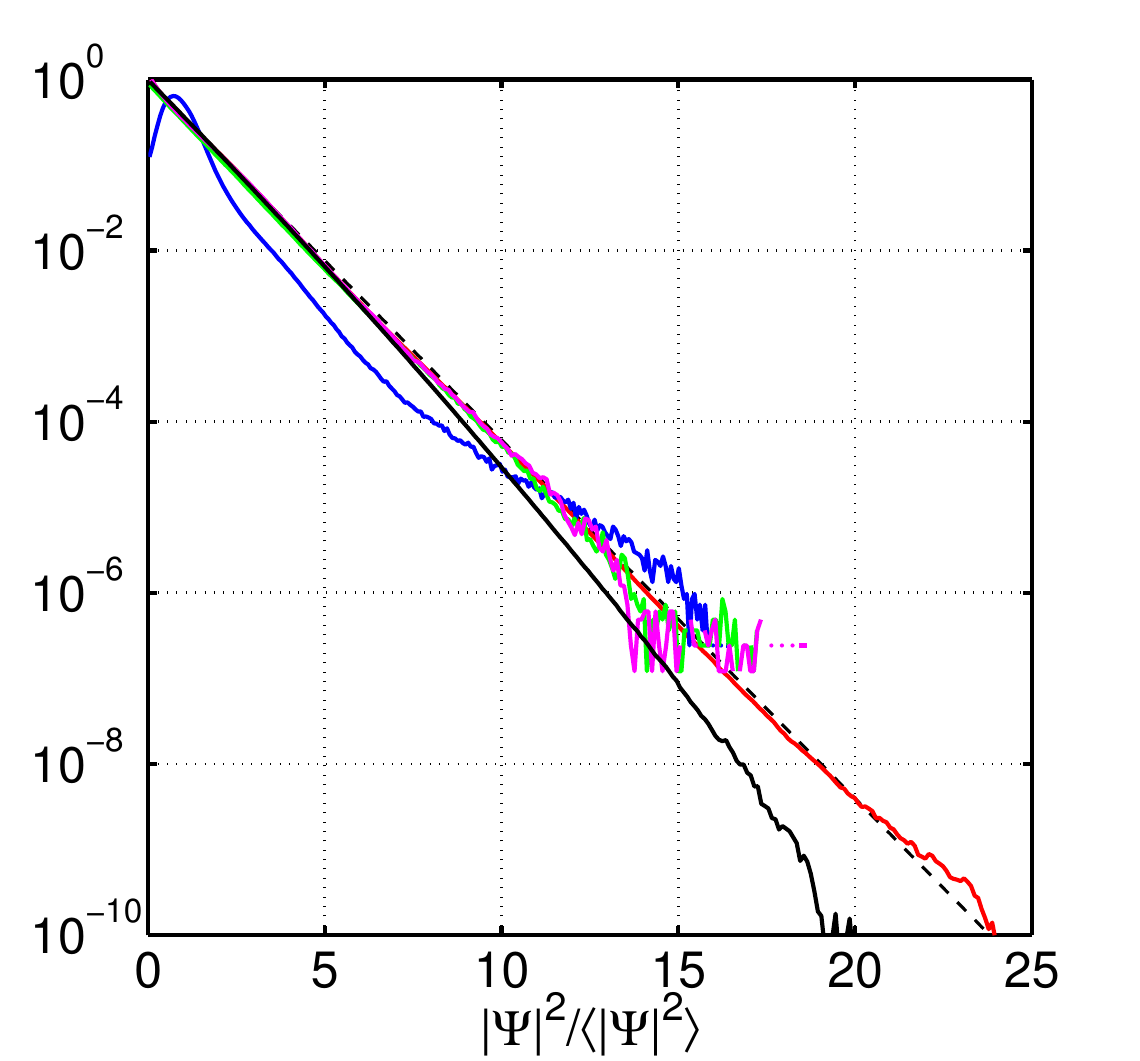}
\caption{\small {\it  (Color on-line) Averaged over ensemble normalized squared amplitudes PDFs in semi-log scale for condensate initial condition for the generalized NLS equation accounting for dumping and pumping terms (\ref{Eq01_1}), $d_{l}=0.0324$, $d_{2p}=0$, $d_{3p}=0.0001$, $p=0.02$. Dashed line is exponential $\sim\exp(-|\Psi|^{2}/2\sigma^{2})$ dependency with $\sigma\approx 1.02$. Blue line corresponds to $t=14$, green - to $t=30$, purple - to $t=65$, red - to time-averaged at $t\in [50, 250]$ PDF, black - to time-averaged PDF of a linear system with the same spectra $I_{k}$ as for the generalized NLS equation accounting for dumping and pumping terms (\ref{Eq01_1}).}}
\end{figure}

\begin{figure}[h] \centering
\includegraphics[width=130pt]{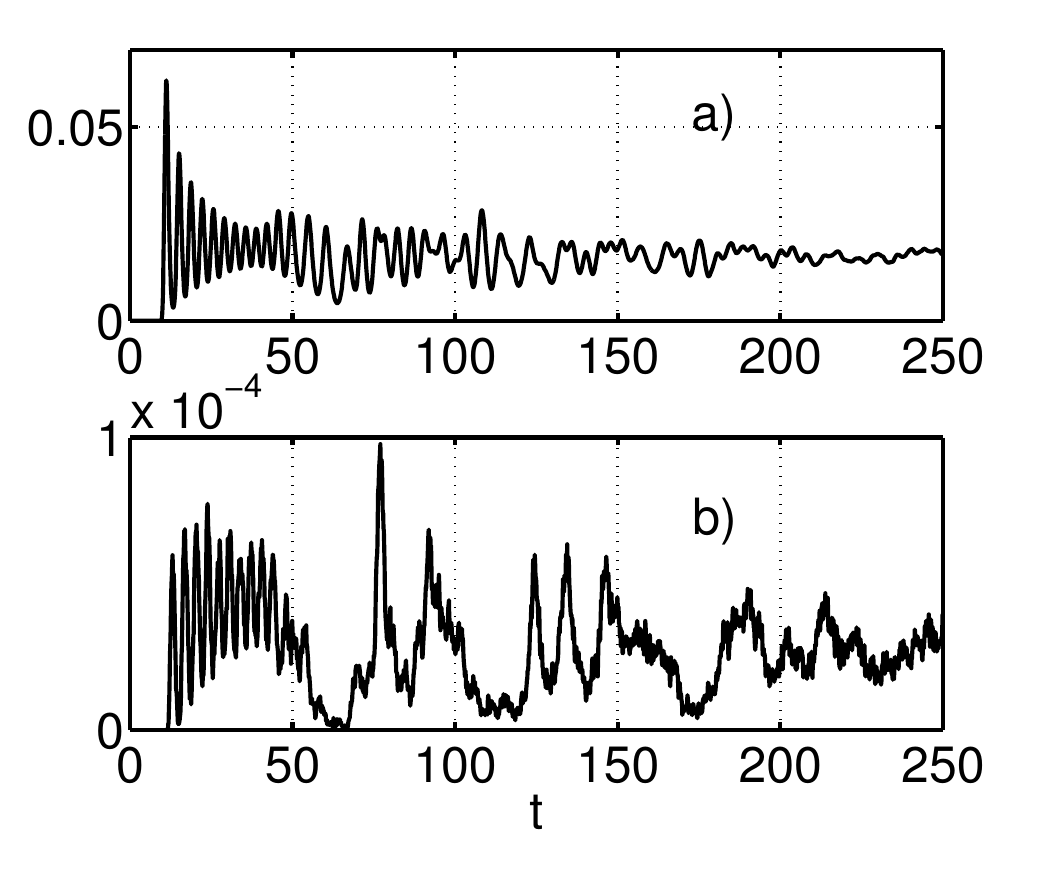}
\includegraphics[width=130pt]{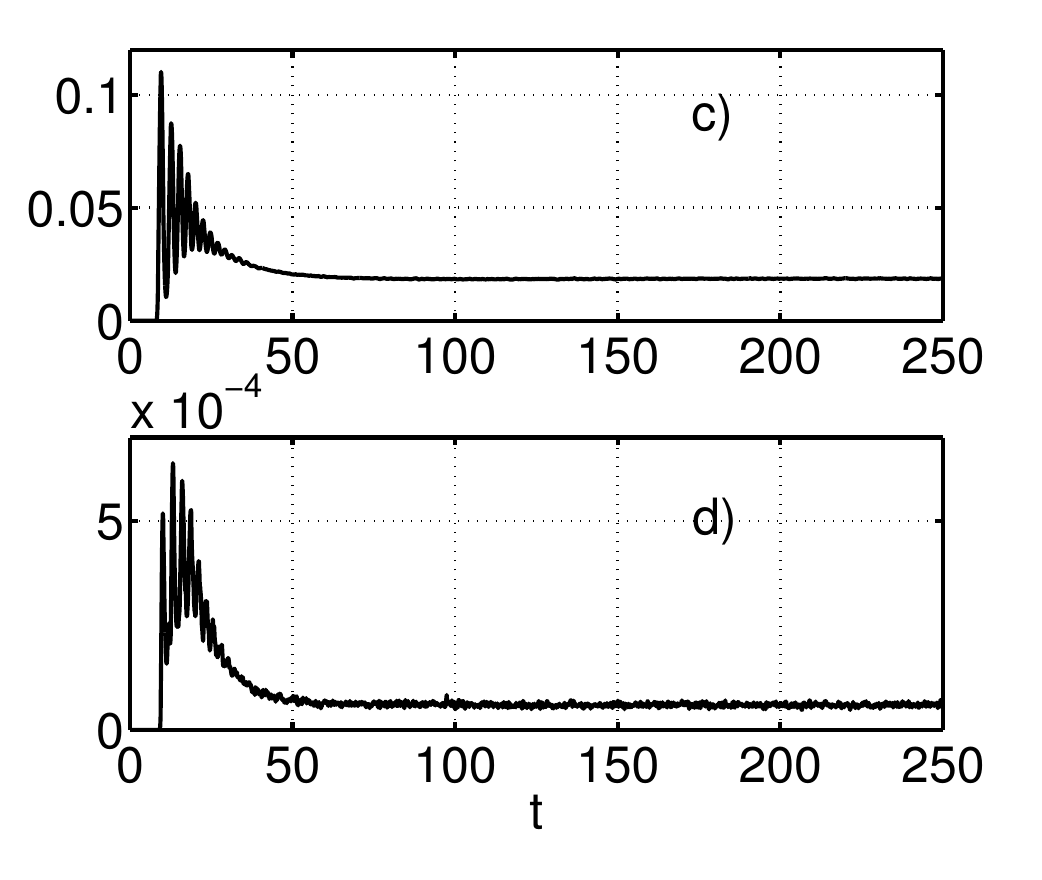}
\caption{\small {\it Frequency of squared amplitudes appearance calculated as averaged over ensemble relative number of points where squared amplitude exceedes $A_{1}^{2}=4$ (a) and (c) and $A_{2}^{2}=10$ (b) and (d) for the classical NLS equation (\ref{Eq01}) (a) and (b) and generalized NLS equation accounting for dumping and pumping terms (\ref{Eq01_1}), $d_{l}=0.0324$, $d_{2p}=0$, $d_{3p}=0.0001$, $p=0.02$, (c) and (d).}}
\end{figure}

\begin{figure}[h] \centering
\includegraphics[width=130pt]{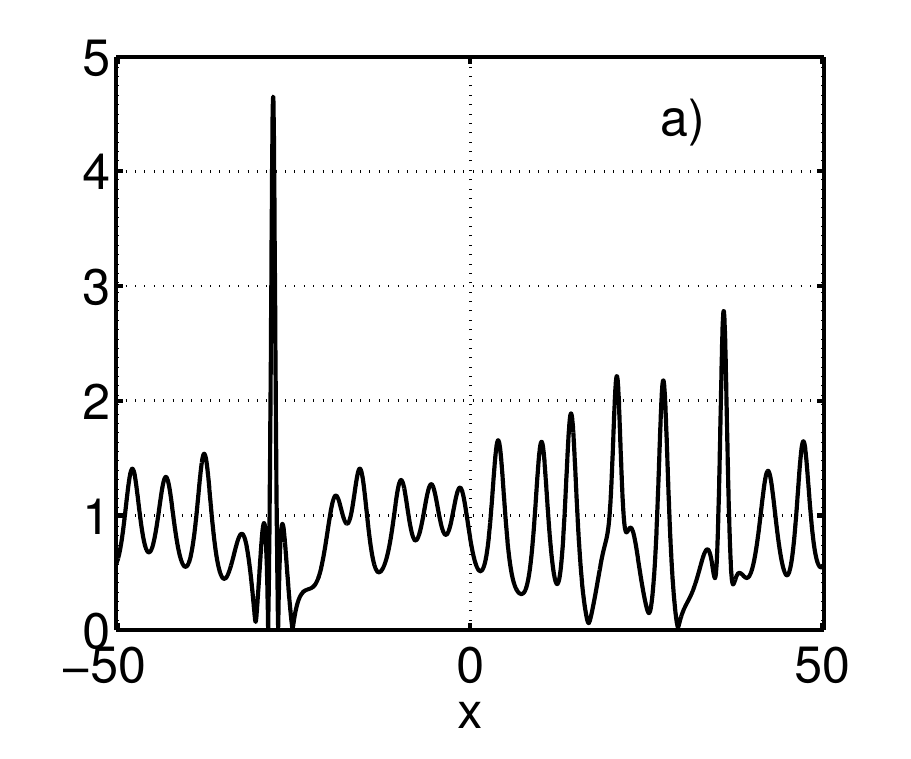}
\includegraphics[width=130pt]{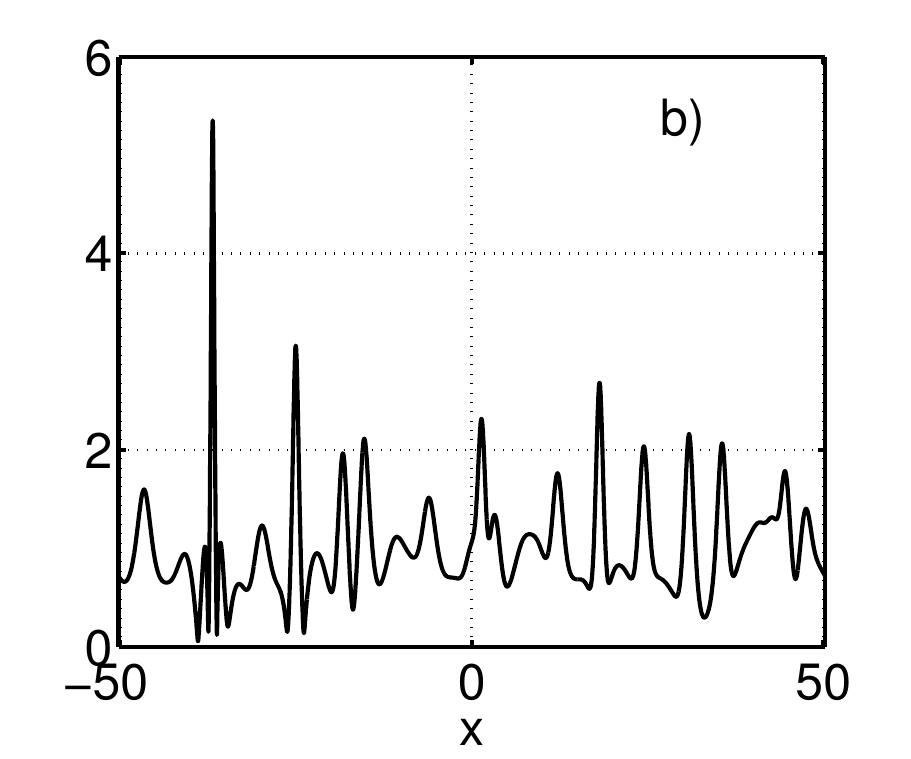}\\
\includegraphics[width=130pt]{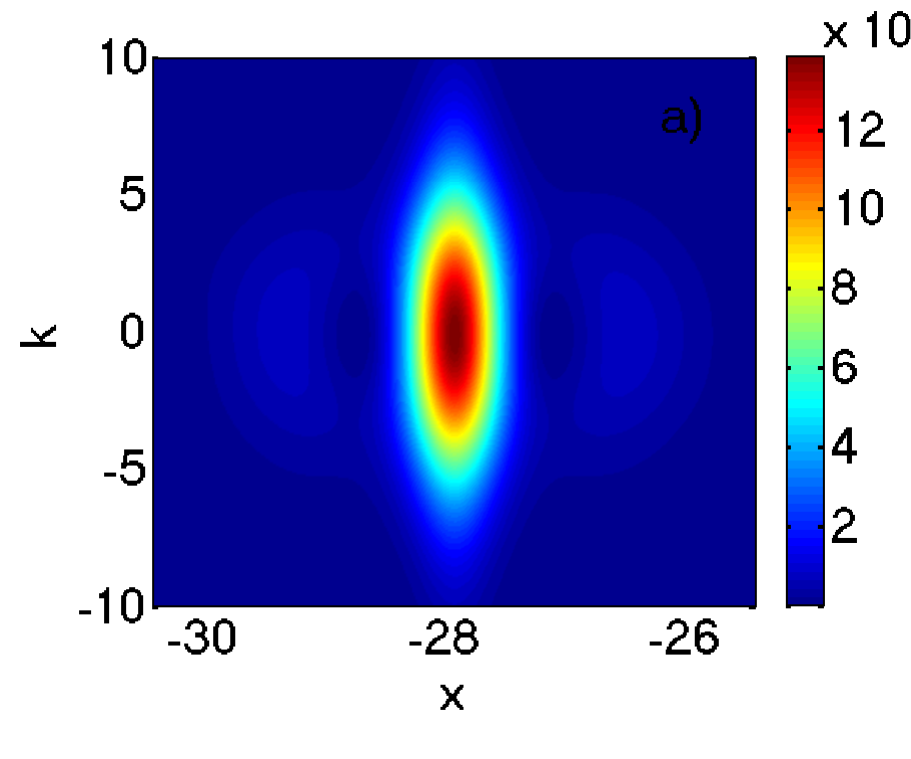}
\includegraphics[width=130pt]{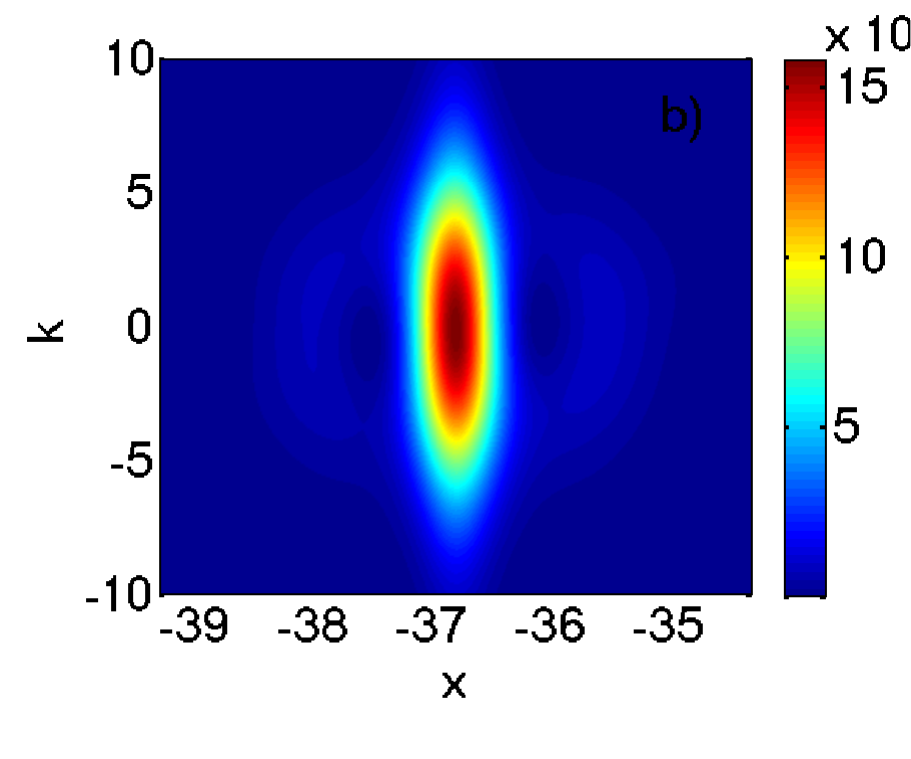}
\caption{\small {\it  (Color on-line) Field distributions $|\Psi|$ and spectrograms of typical large wave events for condensate initial condition for the classical NLS equation (\ref{Eq01}), $|\Psi|_{max}=4.7$, $t=24.9$ (a) and generalized NLS equation accounting for dumping and pumping terms (\ref{Eq01_1}), $|\Psi|_{max}=5.4$, $t=16.1$ (b).}}
\end{figure}

In this paper we choose ensembles of initial distributions in the form of modulationally unstable solutions seeded by noise because of the three main reasons. First, there is a common opinion that optical rogue waves are typically observed during modulation instability development \cite{Solli, Dudley2}. Second, such choice of initial conditions is very convenient since it allows to observe very broad dynamics of the system when the motions of the subsequent initial distributions are very different from each other starting from the nonlinear stage of modulation instability. Third, for small initial noise $\epsilon(x)$ deviations in initial wave action and energy are also small compared to their mean values (unlike for Gaussian-distributed linear waves $\Psi|_{t=0}=C_{0} \int exp(-k^{2}/\Gamma^{2}+i\xi_{k})\,exp(ikx)\,(dk/\sqrt{2\pi})$ for example). The latter allows to study statistical properties for the conservative system (\ref{Eq01}) inside one common class of solutions with fixed integral characteristics 
when, for example, different parts of a PDF corresponding to small, medium and large waves are made up from distributions $\Psi$ with the same fixed wave action and energy.

Starting from the described above initial conditions we observe modulation instability that develops to $t\sim 10$ for Eq. (\ref{Eq01})-(\ref{Eq01_1}) and leads to formation of one-dimensional wave turbulence. In the integrable case (\ref{Eq01}) the turbulence is called integrable and relaxes to one of the infinite possible stationary states. As shown on FIG. 2a,d, during the evolution kinetic energy $H_{d}$ is comparable to the potential one $H_{4}$ for both systems (\ref{Eq01}) and (\ref{Eq01_1}), that means we are working in the regime of solitonic or quasi-solitonic turbulence when solitons play significant role in the turbulent re-distribution of energy inside the system (see \cite{Zakharov1, Zakharov2}). It is interesting to note that for the integrable case (\ref{Eq01}) up to approximately $t\sim 60$ are clearly seen some very regular oscillations of the mean kinetic and potential energy, especially in the region $t\in[10, 50]$. These oscillations persist with increasing of ensemble, their amplitude 
in the region $t\in[10, 30]$ is significantly higher than the corresponding standard deviations, therefore, this might be a sign of some super-correlated dynamics over the whole ensemble of initial distributions $\Psi|_{t=0}=1+\epsilon(x)$, $|\epsilon(x)| \ll 1$. 

At approximately $t\sim 60$ nonlinear lengths kinetic and potential energy for the integrable system (\ref{Eq01}) approach to their initial values, and this corresponds to quasi-periodic dynamics of the classical NLS equation when numerical solution in almost all of the computational domain can again be represented as initial condensate solution seeded by noise $\Psi=1+\xi(x)$, $|\xi(x)| \ll 1$. Beyond $t\sim 60$ nonlinear lengths oscillations of the mean kinetic and potential energy become less regular and almost completely cease after $t\sim 120$ with the simultaneous increase of the corresponding standard deviations that 'sweep' the same ranges of values as before. We think this behavior is connected with the difference in quasi-periodicity points for different initial conditions that leads to gradual mistiming of the solutions with time, since the other possible reason - error of numerical simulations - is extremely small in the region $t\in[0, 60]$ and does not change the class of the solutions that can 
be described at zero time as $\Psi|_{t=0}=1+\epsilon(x)$, $|\epsilon(x)| \ll 1$. As demonstrated on FIG.2a, beyond the initial modulation instability development and except some small regions of quasi-periodicity points the dynamics of the system does not change qualitatively and this allows us to work with additionally averaged over time spectra, spacial correlation functions and the PDFs even when these characteristics still slightly vary with time. Here and below under averaging over time we mean averaging excluding the initial period of time when modulation instability is not fully developed to its nonlinear stage.

For the nonintegrable case (\ref{Eq01_1}) after the first 20-40 nonlinear lengths the system reaches statistically steady state when energy losses due to the dumping terms are compensated by the pumping term and wave action $N$, total energy $H$ as well as kinetic $H_{d}$ and potential $H_{4}$ energy fluctuate near their mean values as shown on FIG. 2b,c,d. Condensate state $\Psi|_{t=0}^{(0)}=1$ is not the solution of Eq. (\ref{Eq01_1}), at small time shifts $t<8$ there is an uncompensated pumping of energy $ip\Psi$ with the subtraction of much smaller nonlinear dumping $id_{2p}|\Psi|^{2}\Psi+id_{3p}|\Psi|^{4}\Psi$ terms. Such energy pumping leads to uniform increase in wave amplitude from $|\Psi|\sim 1$ to $|\Psi|\sim 1.17$ to time $t=8$, this process is accompanied by the modulation instability that becomes noticeable starting from $t>8$ and develops the same way as for the classical NLS equation. With the development of modulation instability linear dumping term $-id_{l}\Psi_{xx}$ being negligible at 
small time shifts $t<8$ significantly increases and pumps the excess of energy out of the system so that for the given dumping and pumping parameters mean wave action, hamiltonian, kinetic and potential energy approach to $t\sim 50$ to the same values as for the classical NLS equation (see FIG. 2). During the modulation instability development we observe the same regular oscillations of kinetic and potential energy as for the integrable case (compare FIG. 2a and 2d) with the exception that modulation instability starts from slightly higher wave amplitudes. These oscillations cease to time shifts $t\sim 30$ and the system approaches to the statistically steady state when spectra, spatial correlation functions, and the PDFs no longer depend on time that allows us to perform additional averaging over time.

Averaged over time and ensemble spectra, spacial correlation functions and the PDFs for Eq. (\ref{Eq01})-(\ref{Eq01_1}) are shown on FIG. 3-6. Thus, for the integrable case (\ref{Eq01}) averaged spectra contains a high peak occupying zeroth harmonic only and then decays at large wavenumbers with almost the same slope in semi-log scale, while spacial correlation function is close to Gaussian at small lengths $|x|<<x_{corr}$ but then decays to a constant level near 0.5 which in its turn is defined by the peak at zeroth harmonic in spectra. Here $x_{corr}$ is the correlation length defined as full width at half maximum of $g(x)$. This peak at zeroth harmonic with varing with time magnitude is always present even when modulation instability is fully developed and the system is far from its initial stage or quasi-periodicity point, it does not depend on noise properties or size of ensemble, and due to this peak spacial correlation functions never decay 
to zero level. We think that the presence of this peak and non-decaying spacial correlation functions might be a sight that in the integrable case (\ref{Eq01}) during the evolution with time condensate never completely disappears. In the nonintegrable case (\ref{Eq01_1}) spectra resembles that for the classical NLS equation (\ref{Eq01}) but is smooth and does not contain peak at zeroth harmonic, the corresponding spacial correlation functions decay to zero level. For both cases (\ref{Eq01})-(\ref{Eq01_1}) spectra in the regions $k=-1$ and $k=1$ contains some nonregular structures (see FIG. 3b and 5b) that correspond to maximum gain of modulation instability at $|k|=1$.

Our results for statistics of wave amplitudes show that for the classical NLS equation (\ref{Eq01}) even despite significantly nonlinear regime $H_{d}\sim H_{4}$ when weak wave turbulence approach does not work tails for the squared amplitudes PDFs are all exponential beyond the initial stage of modulation instability development (blue, green, purple and red lines on FIG. 4). Outside regions of quasi-periodicity (blue and green lines on FIG. 4) PDFs for squared amplitudes are almost entirely exponential, i.e. amplitudes PDFs are almost Rayleigh ones, with some fluctuations in the regions of medium amplitudes. These fluctuations persist with increasing of ensemble, are seen even at large time shifts $t\sim 500-1000$ and might be a sign of some super-correlated dynamics over the whole ensemble of initial distributions. Nevertheless we would like to underline that despite these fluctuations tails at larger amplitudes are all Rayleigh ones, and the time-averaged PDF almost completely coincides with Rayleigh 
distribution (red line on FIG. 4). In the regions of quasi-periodicity PDFs accept severe L-shape form at middle amplitudes but still have Rayleigh tails at larger amplitudes (purple line on FIG. 4). As expected, maximum amplitudes achieved at these time shifts are significantly smaller than outside quasi-periodicity regions. 

For nonintegrable case (\ref{Eq01_1}) beyond the initial stage of modulation instability tails for the squared amplitude PDFs are also exponential (blue, green, purple and red lines on FIG. 6), while fluctuations in the region of medium amplitudes continue up to $t\sim 30$ until oscillations of kinetic and potential energy are present (compare FIG. 2d and 3c). Beyond $t\sim 30$ entire PDFs turn out to be almost indistinguishable from Rayleigh ones. Since in case of Eq. (\ref{Eq01_1}) mean squared wave amplitude $\langle|\Psi|^{2}\rangle=N/\int dx$ depends on time, on FIG. 6 we plot normalized squared amplitude PDFs depending on $|\Psi|^{2}/\langle |\Psi|^{2}\rangle$ that allows us to examine PDFs for different time shifts all on one graph, in the statistically steady state $t>30$ and even before all these PDFs coincide with each other.

Another valuable insight concerning the dynamics of Eq. (\ref{Eq01}) and (\ref{Eq01_1}) might be obtained from the consideration of how frequency of waves amplitude appearance depends on time (see FIG. 7). In the integrable case (\ref{Eq01}) this frequency contains some very regular oscillations with time up to time shifts $t\sim 50$, as shown on FIG. 7a,b. For small waves $|\Psi|^{2}>4$ oscillations nearly cease at much larger times $t\sim 200$ that might be connected to difference in quasi-periodicity points between different realizations inside ensemble. For larger waves $|\Psi|^{2}>10$ significant fluctuations of waves appearance frequency are still seen even at large time shifts. In the nonintegrable case (\ref{Eq01_1}) up to medium time shifts $t\sim 30$ while the oscillations of kinetic and potential energy are present (compare FIG. 2d and FIG. 7c,d) the frequency of waves amplitude appearance also oscillates with time, but then approaches to some constant level with no oscillations noticeable.

Field distributions $|\Psi|$ and spectrograms enlarged at pulse maximums (see \cite{Agafontsev, Treacy}) as well as the motion dynamics for typical large wave events show that for both Eq. (\ref{Eq01}) and (\ref{Eq01_1}) extreme waves composing the tails of the PDFs are collisions of solitons (see FIG. 8).\\


{\bf 2b. Nonlinear Schrodinger equation: cnoidal wave.} \\

In this section we list our results concerning the classical NLS equation (\ref{Eq01}) with initial conditions in the form of cnoidal wave $F(x)$ seeded by stochastic noise $\epsilon(x)$, $\Psi|_{t=0}=F(x)+\epsilon(x)$. Cnoidal wave $F(x)$ is an exact periodic solution of Eq. (\ref{Eq01}) that turns out to be modulationally unstable in case of the focusing four-wave interactions and can be written down as 
$$
F(x)=\sqrt{2Re\wp(\omega_{0})-2Re\wp(x+i\omega_{1})},
$$
where $\wp(z)$ is elliptic Weierstrass function corresponding to half-periods $\omega_{0}$ and $\omega_{1}$ over real and imaginary axis respectively (see \cite{Kuznetsov1}). Cnoidal wave $F(x)$ can also be described as a soliton lattice; half-period over real axis $\omega_{0}$ defines its period over x-axis, $F(x+2\omega_{0})=F(x)$, while half-period over imaginary axis $\omega_{1}$ determines overlapping between solitons. In the current publication we fix real half-period to $\omega_{0}=\pi$, then for relatively small imaginary half-periods $\omega_{1}\sim 1$ cnoidal wave $F(x)$ represents a lattice of almost non-interacting relatively thin and high solitons, while for larger $\omega_{1}$ overlapping between solitons increases and eventually cnoidal wave coincides with the condensate solution $F(x)=1/\sqrt{2}$ as $\omega_{1}\to+\infty$. 

\begin{figure}[h] \centering
\includegraphics[width=130pt]{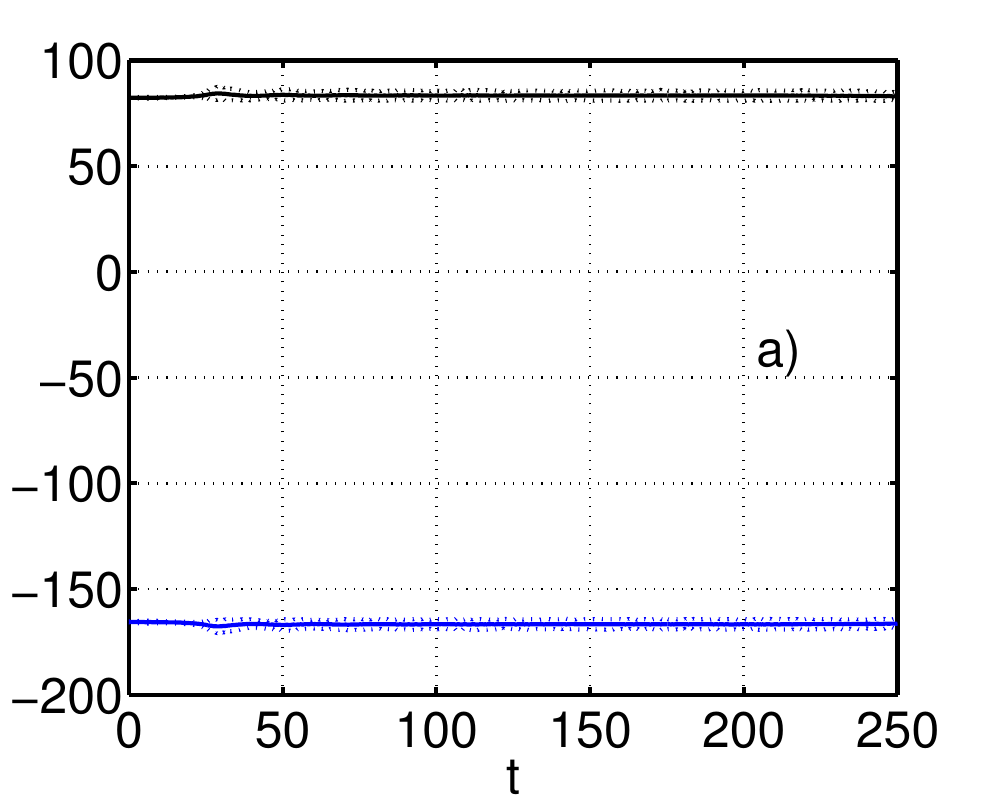}
\includegraphics[width=130pt]{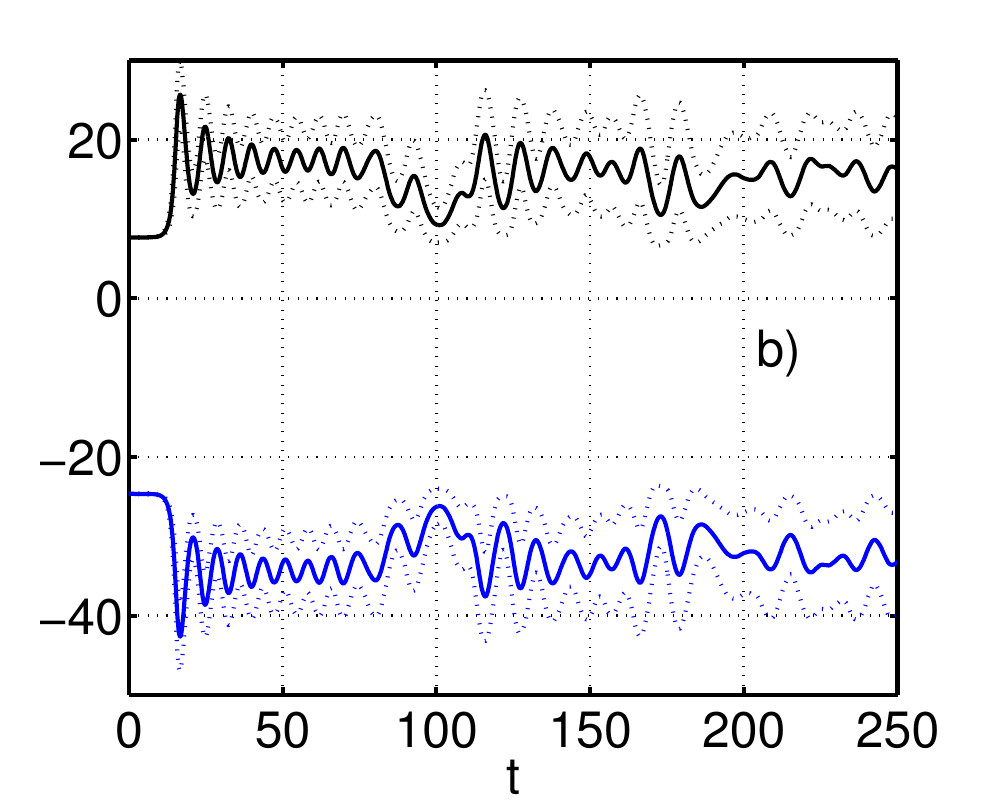}
\caption{\small {\it  (Color on-line) Evolution of averaged over ensemble kinetic $\langle H_{d}\rangle$ (black) and potential $\langle H_{4}\rangle$ (blue) energy for the classical NLS equation (\ref{Eq01}) for cnoidal wave initial conditions with parameters $\omega_{0}=\pi$, $\omega_{1}=1$ (a) and $\omega_{0}=\pi$, $\omega_{1}=2$ (b). Solid lines - mean over ensemble values, dashed lines - borders for the corresponding standard deviations.}}
\end{figure}

\begin{figure}[h] \centering
\includegraphics[width=130pt]{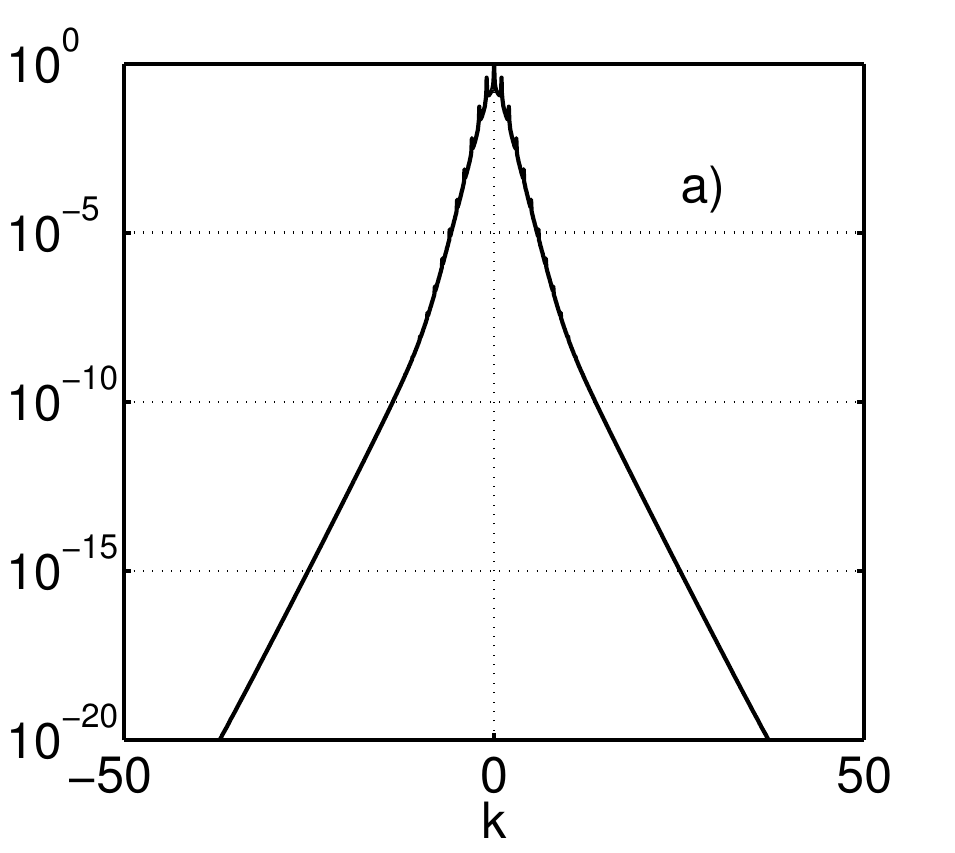}
\includegraphics[width=130pt]{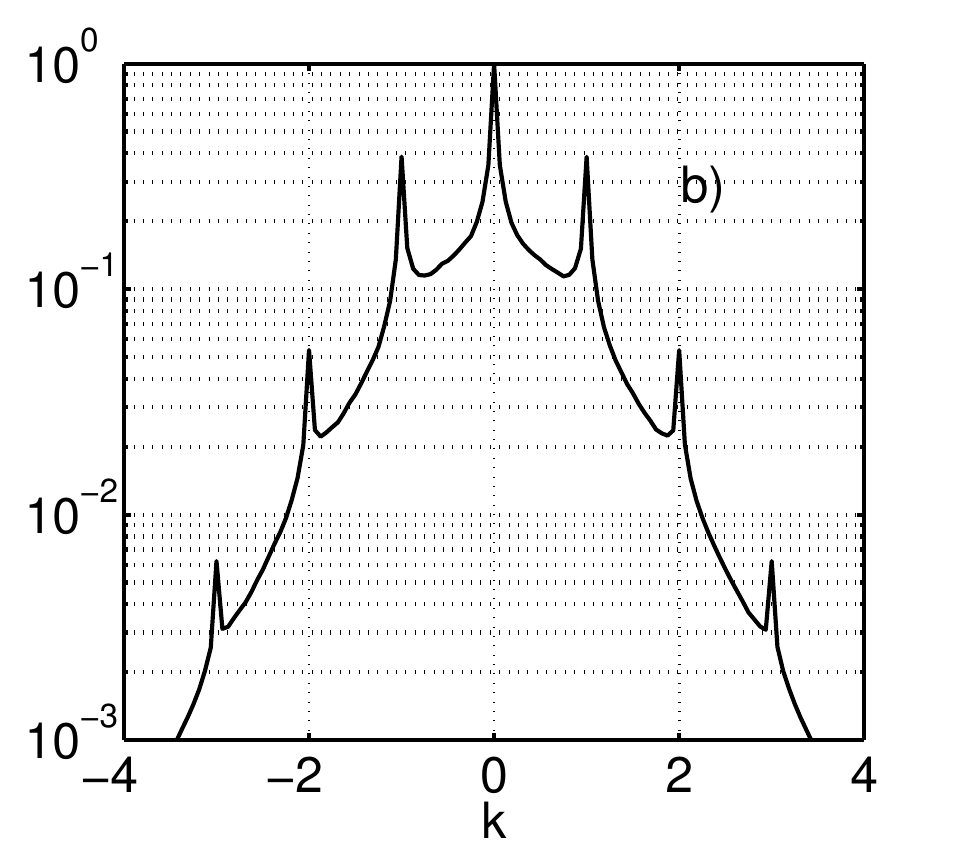}
\includegraphics[width=130pt]{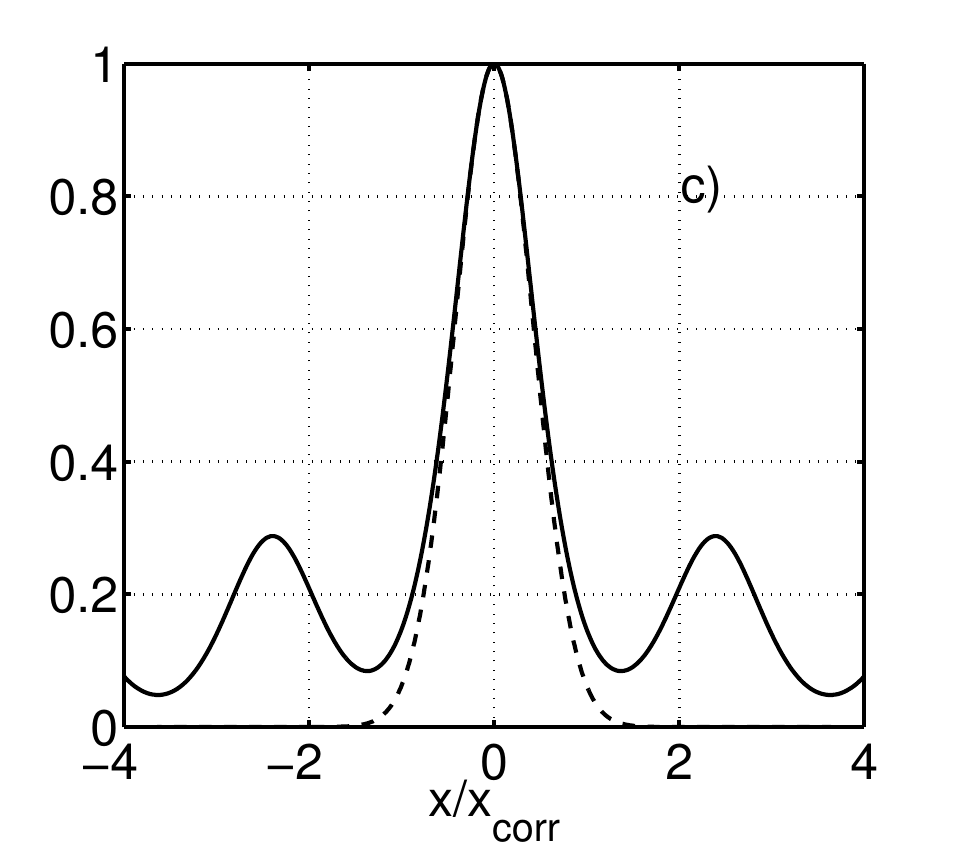}
\caption{\small {\it  Averaged over ensemble and time $t\in [20, 250]$ normalized spectra $I_{k}/\max I_{k}$ in full scale (a) and enlarged at center region (b) as well spatial correlation function $g(x/x_{corr})/g(0)$ for the classical NLS equation (\ref{Eq01}) for cnoidal wave ($\omega_{0}=\pi$, $\omega_{1}=1$) initial condition. For graph (c) dashed line is Gaussian distribution.}}
\end{figure}

\begin{figure}[h] \centering
\includegraphics[width=320pt]{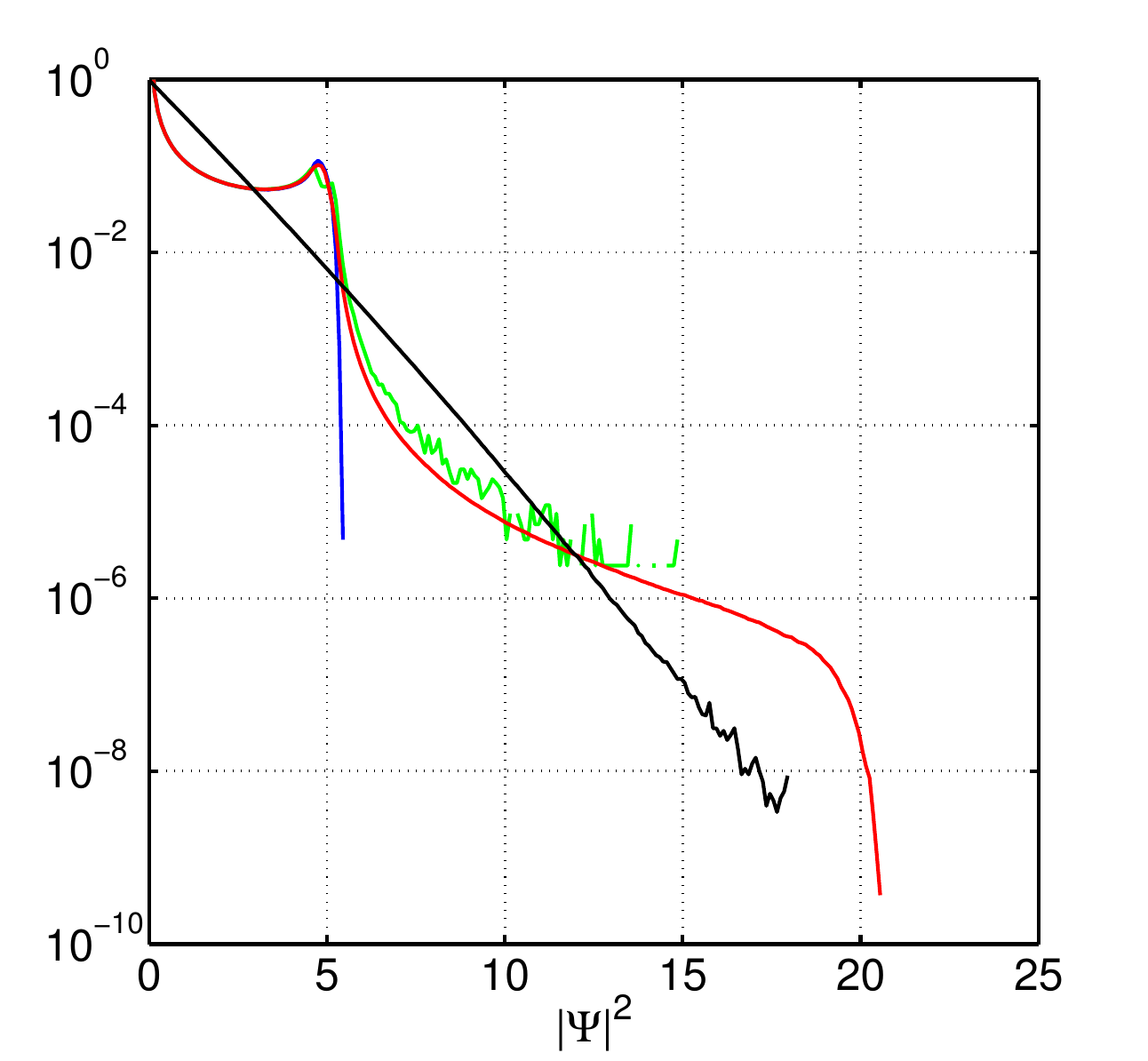}
\caption{\small {\it (Color on-line) Averaged over ensemble normalized squared amplitudes PDFs in semi-log scale for the classical NLS equation (\ref{Eq01}) for cnoidal wave ($\omega_{0}=\pi$, $\omega_{1}=1$) initial condition. Blue line corresponds to $t=20$, green - to $t=30$, red - to time-averaged at $t\in [20, 250]$ PDF, black - to time-averaged PDF of a linear system with the same spectra $I_{k}$ as for the classical NLS equation (\ref{Eq01}).}}
\end{figure}

Evolution of the averaged over ensemble kinetic and potential energy for two ensembles of initial distributions based on cnoidal waves with parameters $\omega_{0}=\pi$, $\omega_{1}=1$ and $\omega_{0}=\pi$, $\omega_{1}=2$ is shown on FIG. 9a and 9b respectively. When overlapping between solitons is small enough wave field $\Psi$ stays close to the initial cnoidal wave $F(x)$ almost all the time during its evolution with the exception of very rare two-solitons collisions that results in very small oscillations of the mean kinetic and potential energy near their corresponding values for pure cnoidal wave (see FIG. 9a). Dynamics of the system becomes more rich with three-, four- and so on soliton collisions present as the overlapping between solitons increases, so that evolution of kinetic and potential energy resembles that for the condensate case with very similar regular oscillations clearly visible at moderate time shifts up to $t\sim 100$ (FIG. 9b). As for the condensate case due to gradual mistiming of the 
solutions owing to difference in quasi-periodicity points beyond $t\sim 100$, these oscillations gradually cease with the simultaneous increase of the corresponding standard deviations that 'sweep' the same ranges of values as before when the oscillations were present.

\begin{figure}[h] \centering
\includegraphics[width=130pt]{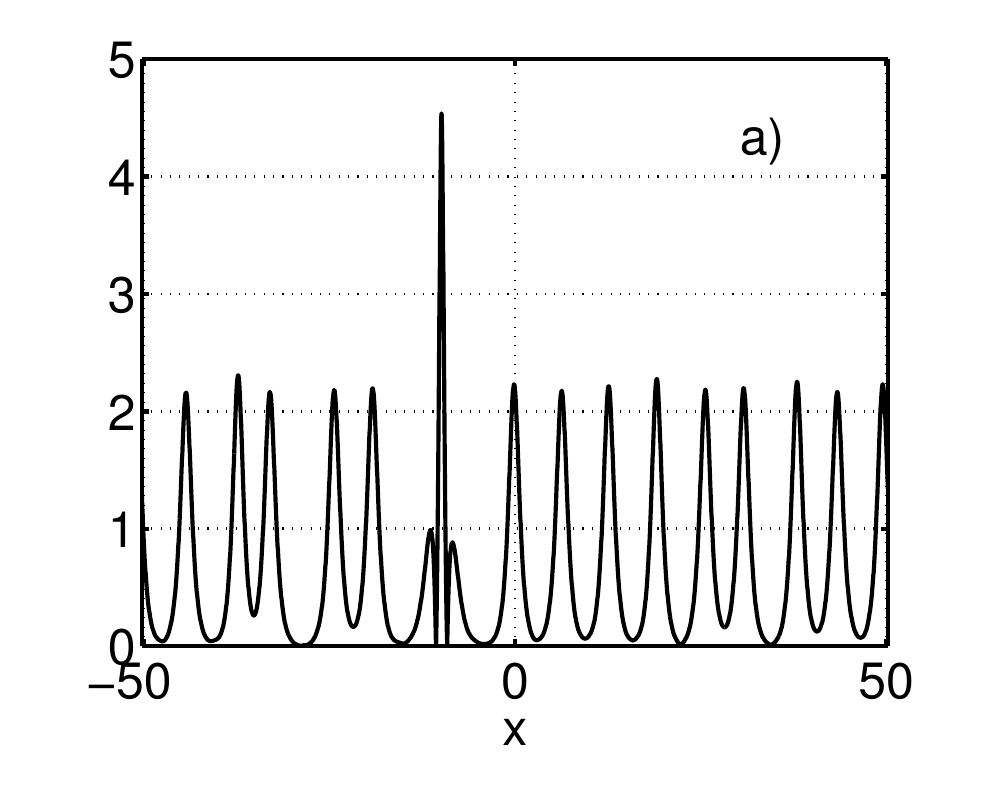}
\includegraphics[width=130pt]{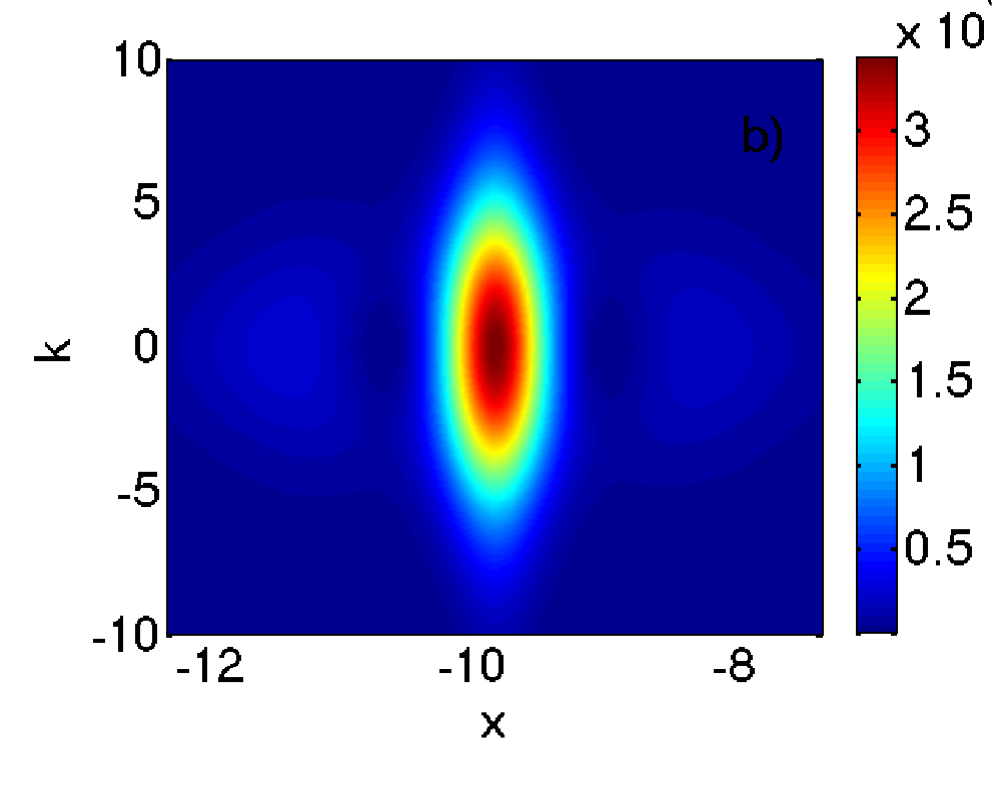}
\includegraphics[width=130pt]{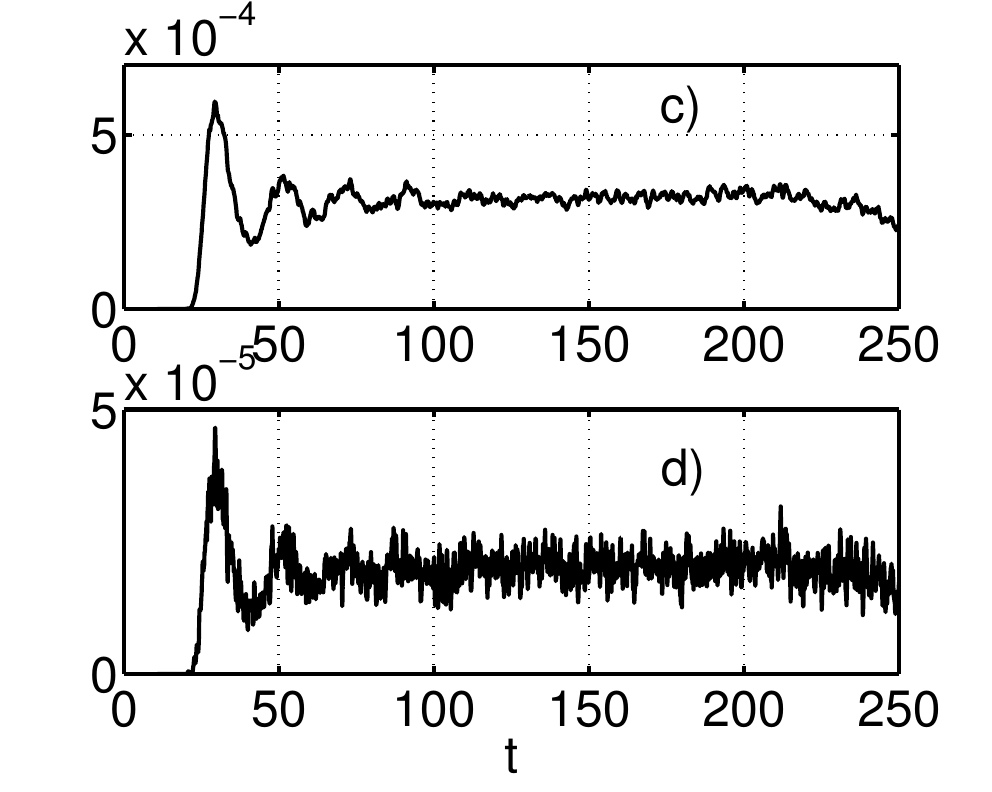}
\caption{\small {\it  (Color on-line) Field distribution $|\Psi|$ (a) and spectrogram (b) for a typical large wave event, $|\Psi|_{max}=4.5$, $t=82.6$, frequency of squared amplitudes appearance calculated as averaged over ensemble relative number of points where squared amplitude exceedes $A_{1}^{2}=6$ (c) and $A_{2}^{2}=10$ (d) for the classical NLS equation (\ref{Eq01}) with cnoidal wave initial condition, $\omega_{0}=\pi$, $\omega_{1}=1$.}}
\end{figure}

\begin{figure}[h] \centering
\includegraphics[width=130pt]{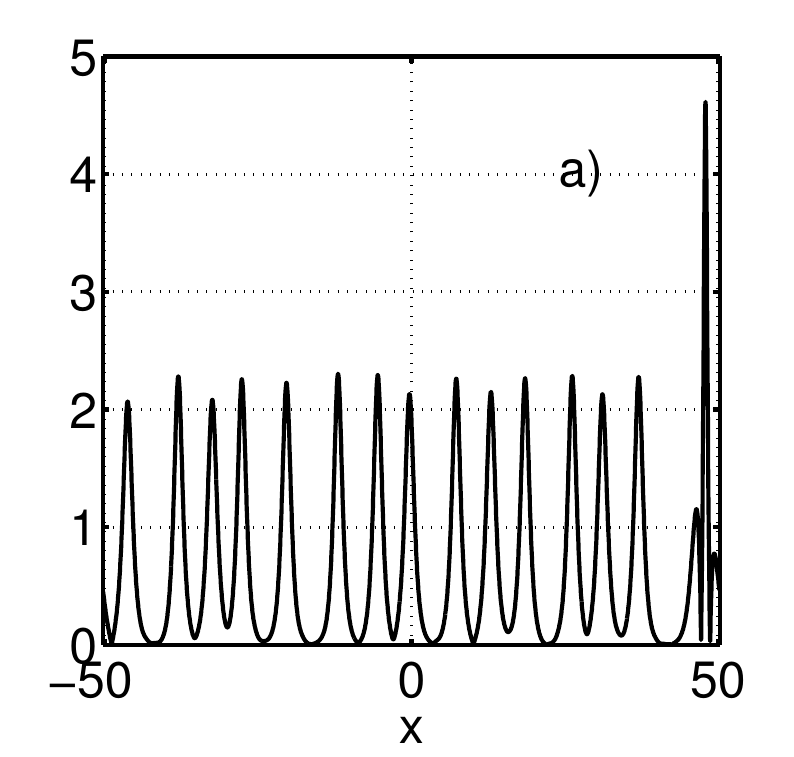}
\includegraphics[width=130pt]{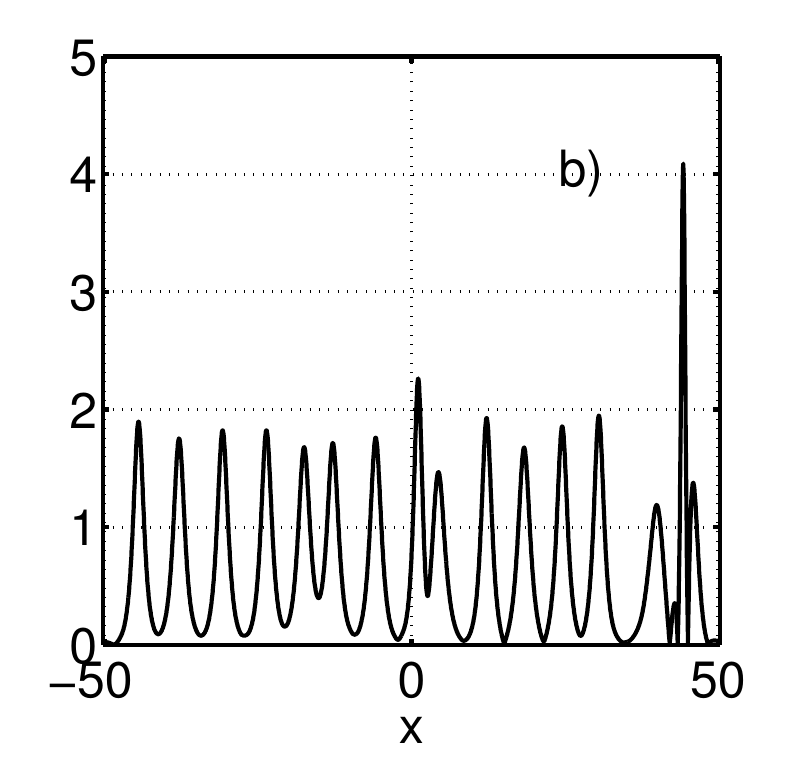}
\includegraphics[width=130pt]{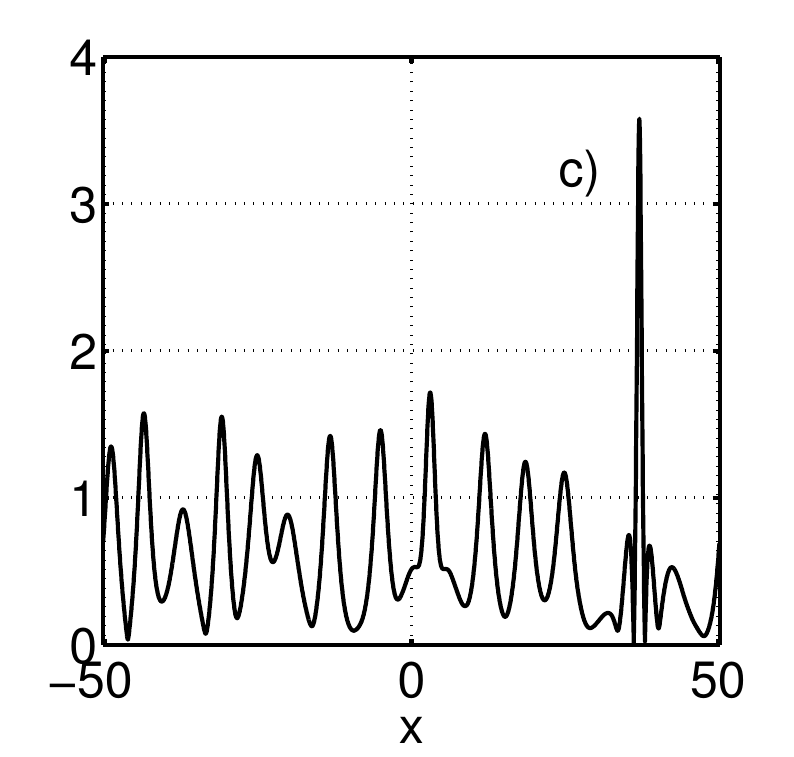}
\includegraphics[width=130pt]{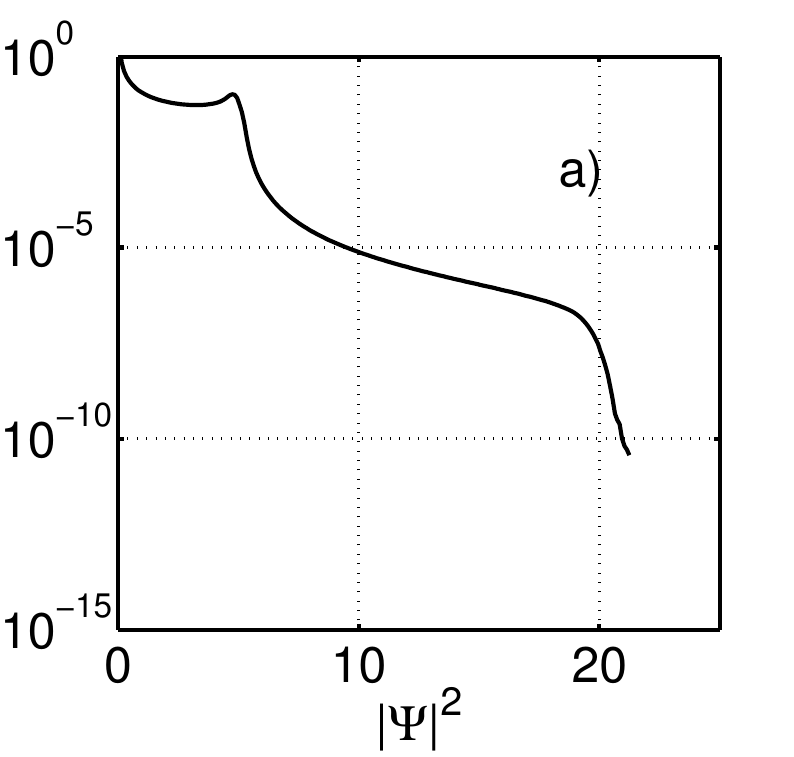}
\includegraphics[width=130pt]{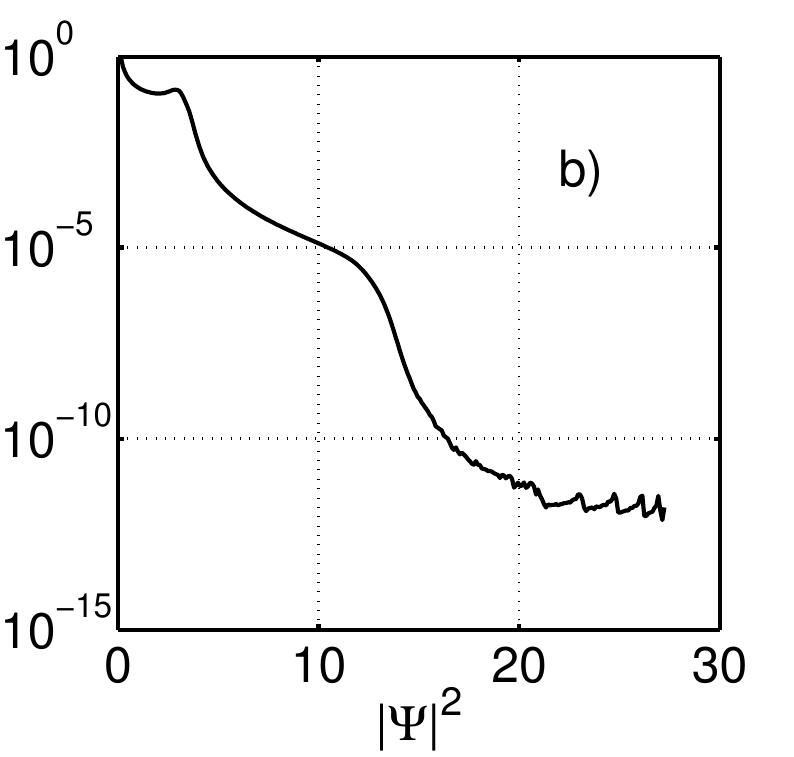}
\includegraphics[width=130pt]{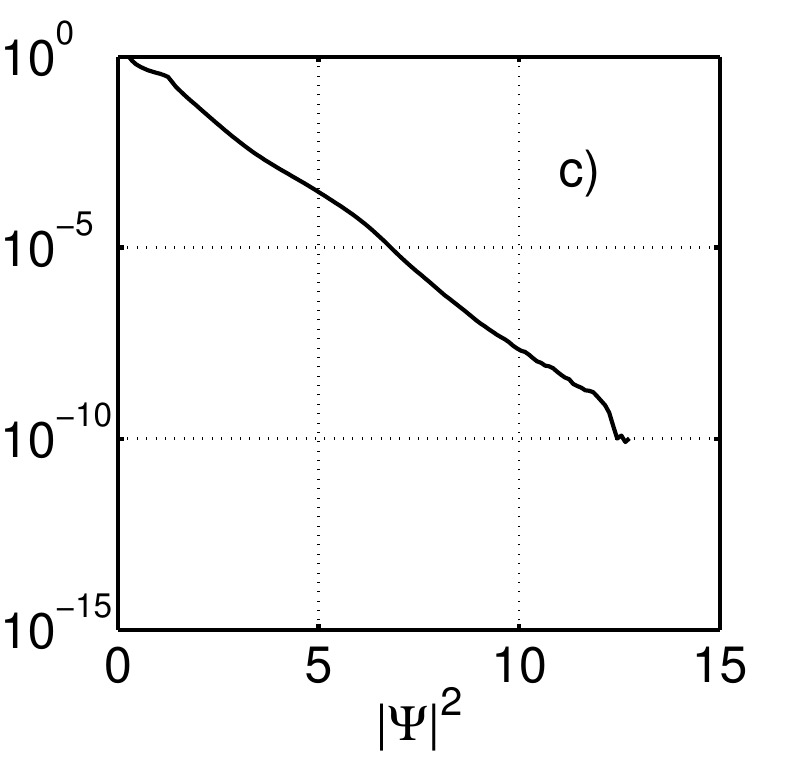}
\caption{{Typical rogue wave events and averaged over time and ensemble squared amplitudes PDFs for cnoidal waves with parameters $\omega_{0}=\pi$, $\omega_{1}=1$ (a), $\omega_{0}=\pi$, $\omega_{1}=1.25$ (b) and $\omega_{0}=\pi$, $\omega_{1}=2$ (c).}}
\end{figure}

As shown on FIG. 10, averaged over ensemble and time spectra contains equidistant peaks at integer wavenumbers survived from those of the initial cnoidal wave $F(x)$, some transition region at moderate wavenumbers and then exponentially decaying tails at large wavenumbers. For significant overlapping of solitons there is also noticeable peak at zeroth harmonic as for the condensate case, its magnitude gradually increases with overlapping approaching the same values as for the condensate initial condition in the limit $\omega_{1}\to+\infty$. Accordingly, spacial correlation functions being close to Gaussian for $|x|<x_{corr}$ then decay with pronounced oscillations to some noticeable nonzero level for significant overlapping of solitons, this level increases with overlapping to $\sim 0.5$ in the limit $\omega_{1}\to+\infty$. PDFs for cnoidal waves with small overlapping between solitons ($\omega_{0}=\pi$, $\omega_{1}=1$ for example) are strongly non-Rayleigh ones and consist of two curved parts (see FIG. 11). 
The first one, corresponding to small amplitudes, represents amplitudes distribution for the initial cnoidal wave $F(x)$ since $|\Psi|$ stays close to $F(x)$ almost all the time during its evolution. The second part of the PDFs, corresponding to high waves, represents appearance of rare extreme events. Field distribution $|\Psi|$ and spectrogram enlarged at pulse maximum for a typical large wave event shown on FIG. 12a,b and also the motion dynamics demonstrate that these extreme events are two-soliton collisions. The corresponding frequency of wave amplitudes appearance (see FIG. 12c,d) after the first few significant oscillations during modulation instability development then fluctuates near its mean value, these fluctuations become more pronounced as the overlapping between solitons increases.

It is interesting to note that when the overlapping between solitons inside the initial cnoidal wave is small and the collisions of solitons are very rare events, the regions corresponding to two-, three- and so on soliton collisions are easily read on the PDFs as demonstrated on FIG. 13. Thus, in case of cnoidal wave with parameters $\omega_{0}=\pi$, $\omega_{1}=1$ overlapping between solitons is small, dynamics consists mainly of two-solitons collisions and there are two curved parts seen on the time-averaged PDF corresponding to amplitudes distribution of the initial unperturbed cnoidal wave and two-soliton collisions respectively (FIG. 13a). Increasing of imaginary half-period to $\omega_{1}=1.25$ results in significantly more frequent soliton interactions when not only two- but also three-soliton collisions are present, and the corresponding PDF consists of three curved parts (FIG. 13b): amplitudes distribution of the initial cnoidal wave, two- and three- soliton collisions respectively. Further 
increase of imaginary half-period to $\omega_{1}=2$ leads to even more rich dynamics of the system when many-soliton collisions are present. Since such collisions become quite frequent, borders between two-, three- and so on soliton collisions are washed out and the corresponding PDF turns out to be almost Rayleigh one (FIG. 13c).\\


{\bf 3. Generalized Nonlinear Schrodinger equation accounting for six-wave interactions, dumping and pumping terms.} \\

The main purpose of this section is the the investigation of the influence of higher nonlinearity on the statistics of large waves, namely the influence of six-wave interactions that naturally appear as a next term beyond the classical NLS equation in perturbation theory expansion. Here we limit ourselves with only focusing six-wave interactions because this is the most interesting case: it is well known that under certain circumstances addition of focusing six-wave interactions to the classical NLS equation (\ref{Eq01}) results in generation of blow-up collapses in a finite time. Indeed, collisions of waves lead to appearance of high amplitudes for which four-wave interactions are small compared to six-wave interactions and the resulting equations of motion can be approximated by the quintic NLS equation. On the other hand, quintic NLS equation is the well-known critical model $sd=4$, where $s$ is the power of nonlinearity $|\Psi|^{s}\Psi$ and $d$ is spacial dimension, that may develop singularity in the 
form of infinite amplitude in a finite time, or wave collapse \cite{Zakharov3, Zakharov4}. However in the real physical systems with the growth of wave amplitude the other interactions will become significant that will eventually lead to regularization of such collapses; these interactions might be defocusing next-order nonlinear terms (eight-wave interactions for example) that stop amplitude growth starting from the amplitude when these interactions exceed six-wave interactions, or there might appear dumping interactions. 

In this paper we consider regularization with dumping terms only because it turns out that regularizations the help of conservative defocusing next-order nonlinear terms have at least two major drawbacks. First, for such systems we observe non-trivial results at time-shifts where all interactions (dispersion, four-, six- and so on waves scattering) become comparable with each other and therefore the corresponding dynamics is qualitatively different from that of the classical NLS equation. Second, under certain circumstances there exists a statistical attractor in such systems in the form of one large-scale coherent soliton containing all the potential energy and immersed in the field of small perturbations that keeps all the necessary information for time-reversal $t\rightarrow -t$ (see \cite{Jordan} and also \cite{Zakharov1, Zakharov2}). Parameters of the statistical attractor and even its existence are determined by the integrals of motion while the values of the latter ones depend on size of the 
computational domain. 	Accordingly, there is a possibility to obtain significantly different results by simply doubling the computational domain area that we find not physically relevant. 

Therefore in this section we perform regularization of collapses with the help of the same dumping terms as in the section 2a, namely linear dissipation and two- and three-photon absorption, and in order to balance the system we use the same deterministic forcing term $ip\Psi$:
\begin{eqnarray}\label{Eq03}
& i\Psi_t +(1-id_{l})\Psi_{xx}-\Psi+(1+id_{2p})|\Psi|^2 \Psi +(\alpha+id_{3p}) |\Psi|^4 \Psi = ip\Psi,\\ 
& d_{l},d_{2p},d_{3p},p>0, \quad \alpha,d_{l},d_{2p},d_{3p},p \ll 1. \nonumber
\end{eqnarray}
During regularization of collapses generated by six-wave interactions $\alpha|\Psi|^4 \Psi$ nonlinear dissipation prevents formation of waves with too high amplitudes, while linear dissipation prevents appearance of too high gradients. We checked with the same methodology as described in \cite{Lushnikov} that in case of system (\ref{Eq03}) such dumping terms manifest themselves mainly during regularization of collapses, while the forcing term constantly pumps energy into the system and thus balances it. Eq. (\ref{Eq03}) belongs to the class of dissipative equations and wave action $N$, momentum $P$ and energy $H=H_{d}+H_{4}+H_{6}$ where $H_{6}=-(1/3)\int\alpha|\Psi|^6 dx$ become functions of time. In the framework of Eq. (\ref{Eq03}) there is no independent on six-wave interactions coefficient and dumping and pumping parameters universal spatio-temporal dynamics for collapses saturation that significantly obstructs any possible theoretical investigation (compare to \cite{Lushnikov}).

\begin{figure}[h] \centering
\includegraphics[width=130pt]{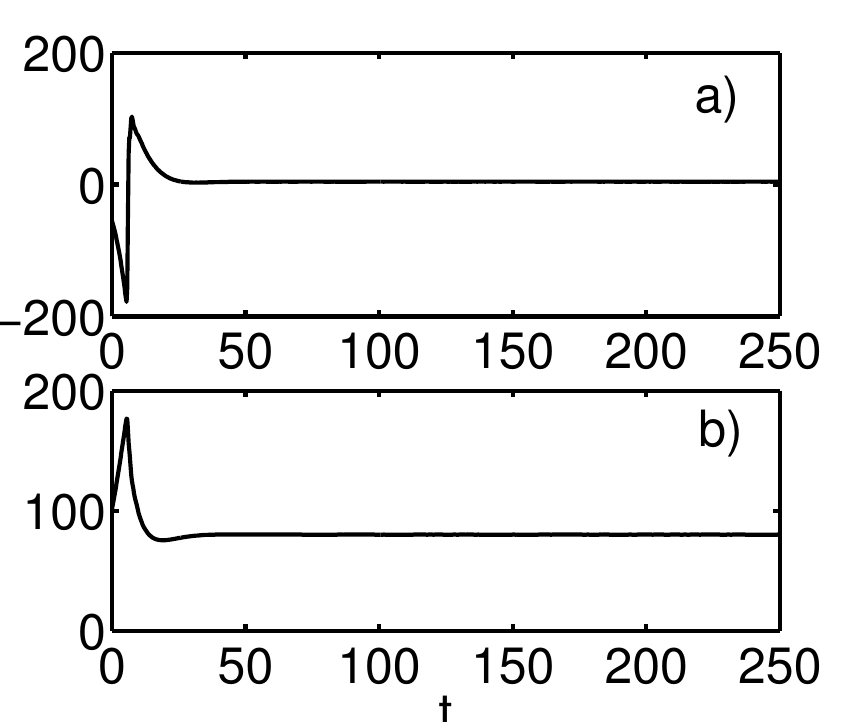}
\includegraphics[width=130pt]{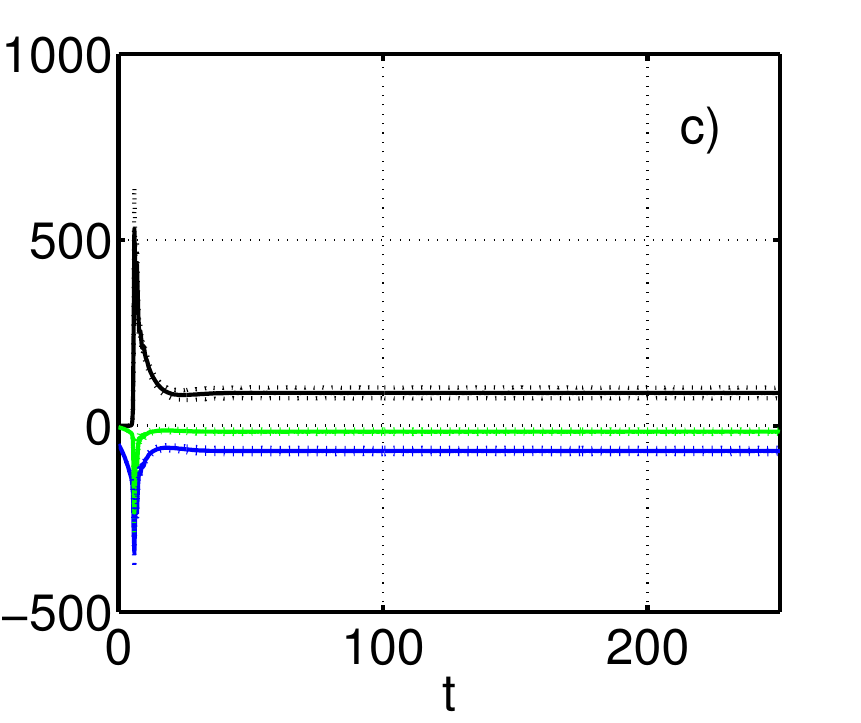}
\caption{\small {\it  (Color on-line) Evolution of averaged over ensemble (a) total energy $\langle H_{d}+H_{4}+H_{6}\rangle$, (b) wave action $\langle N\rangle$ and (c) kinetic energy $\langle H_{d}\rangle$ (black), four- $\langle H_{4}\rangle$ (blue) and six-wave interactions energy $\langle H_{6}\rangle$ (green) for generalized NLS equation accounting for six-wave interactions, dumping and pumping terms (\ref{Eq03}), $\alpha=0.128$, $d_{l}=0.04$, $d_{2p}=0.02$, $d_{3p}=0.0004$, $p=0.05$. Solid lines - mean over ensemble values, dashed lines - borders for the corresponding standard deviations.}}
\end{figure}

\begin{figure}[h] \centering
\includegraphics[width=130pt]{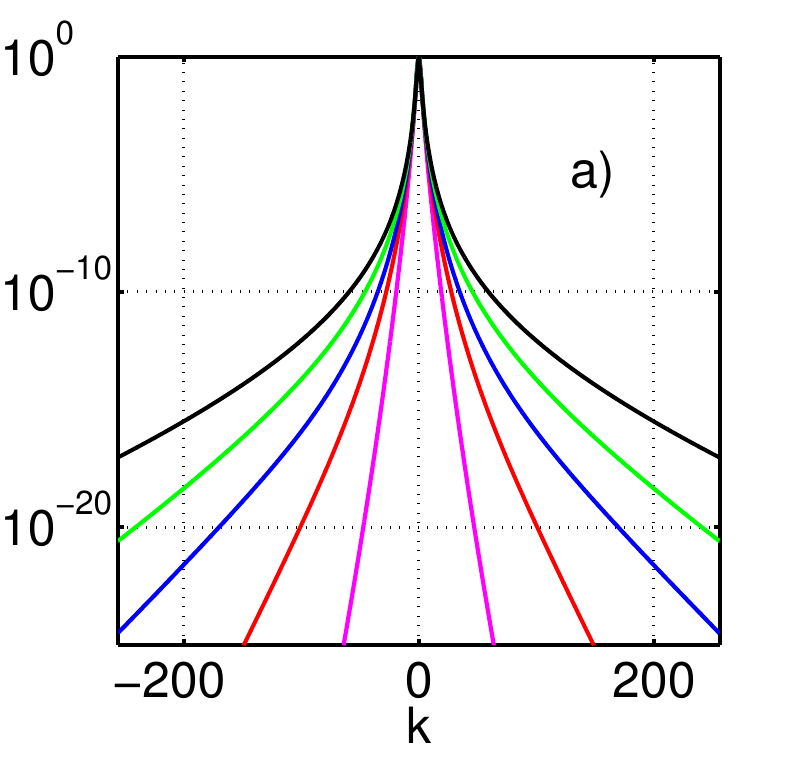}
\includegraphics[width=130pt]{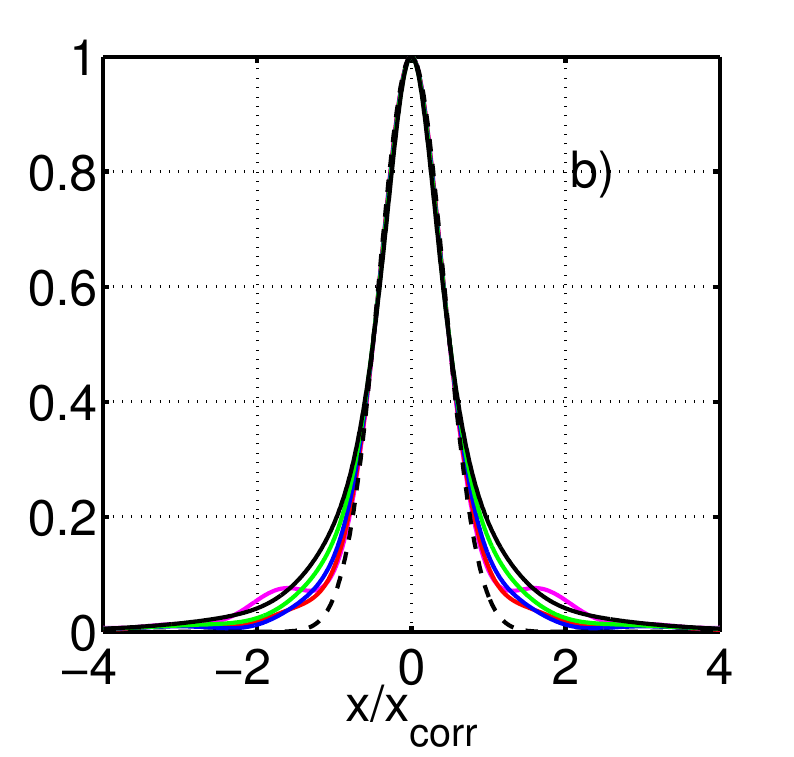}
\includegraphics[width=130pt]{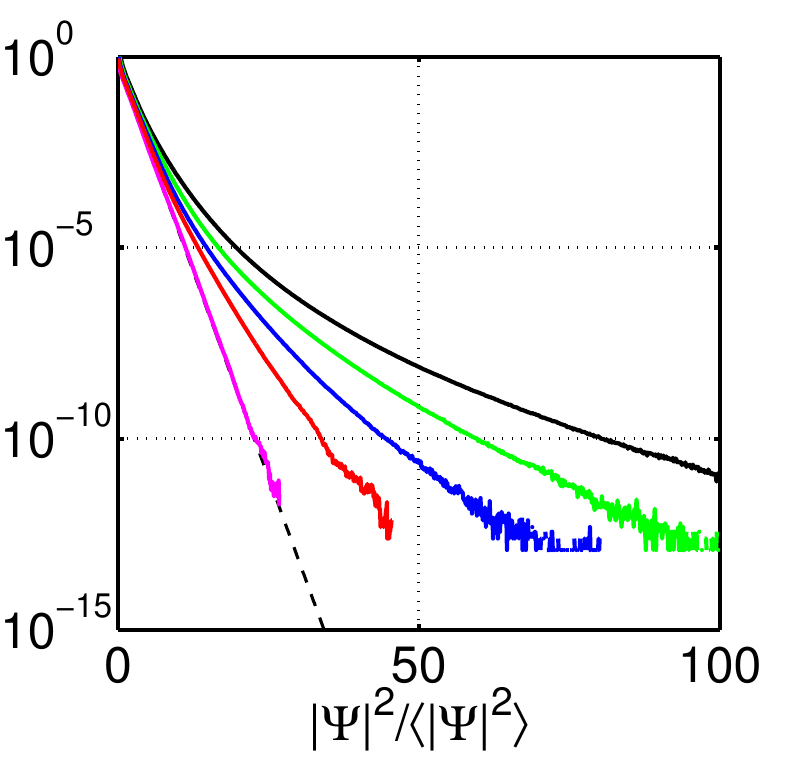}
\caption{\small {\it  (Color on-line) Solid lines are (a) spectra $I_{k}$, (b) normalized spatial correlation functions $g(x/x_{corr})/g(0)$ and (c) normalized squared amplitudes PDFs for generalized NLS equation accounting for six-wave interactions, dumping and pumping terms (\ref{Eq03}) with fixed $d_{l}=0.04$, $d_{2p}=0$, $d_{3p}=0.0004$, $p=0.05$ and $\alpha=0$ (purple), $\alpha=0.032$ (red), $\alpha=0.064$ (blue), $\alpha=0.128$ (green) and $\alpha=0.256$ (black) in the statistically steady states. Dashed line for graphs (b) is Gaussian distribution while for graphs (c) is exponential dependency $\exp(-|\Psi|^{2}/2\sigma^{2})$, $\sigma\approx 0.71$.}}
\end{figure}

\begin{figure}[h] \centering
\includegraphics[width=130pt]{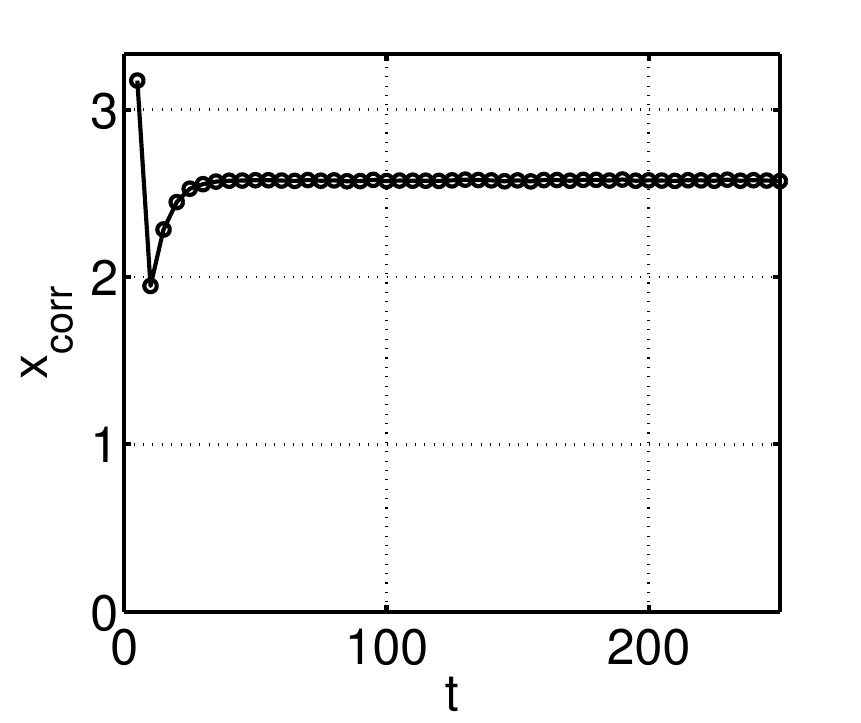}
\caption{\small {\it  Correlation length $x_{corr}$ defined as full width at half maximum of spatial correlation function $g(x)$ depending on time $t$ for generalized NLS equation accounting for six-wave interactions, dumping and pumping terms (\ref{Eq03}), $\alpha=0.256$, $d_{l}=0.04$, $d_{2p}=0$, $d_{3p}=0.0004$, $p=0.05$.}}
\end{figure}

\begin{figure}[h] \centering
\includegraphics[width=130pt]{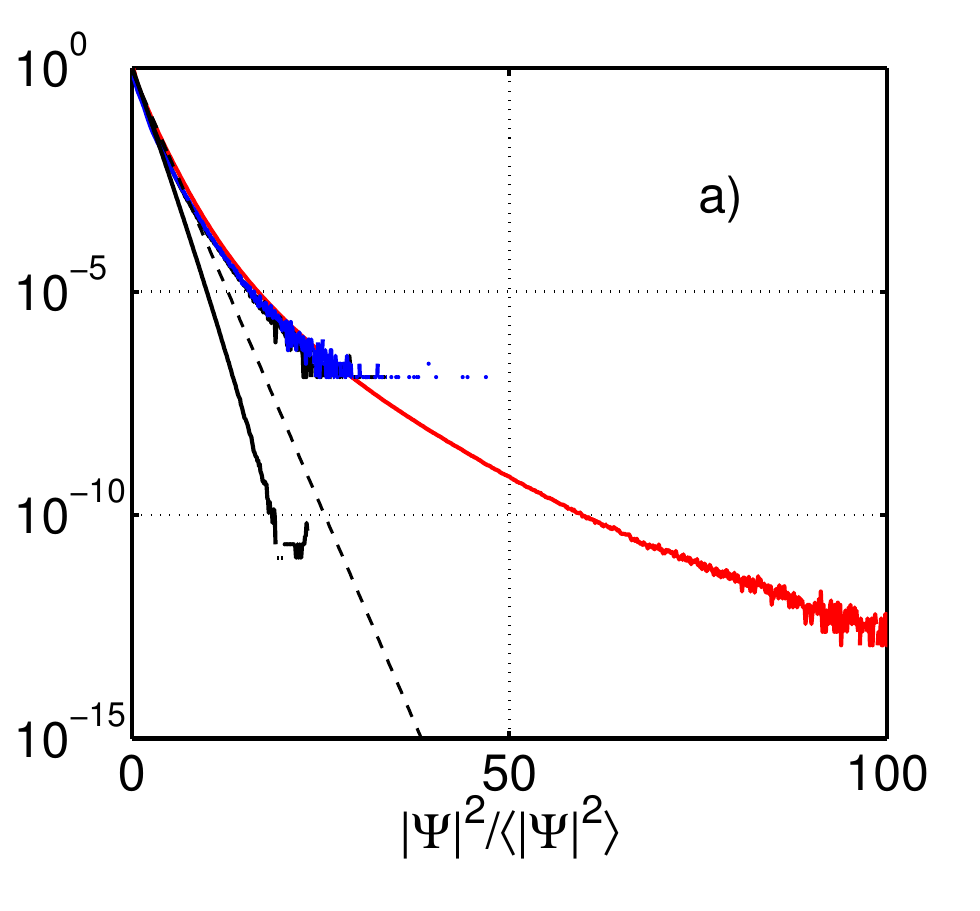}
\includegraphics[width=130pt]{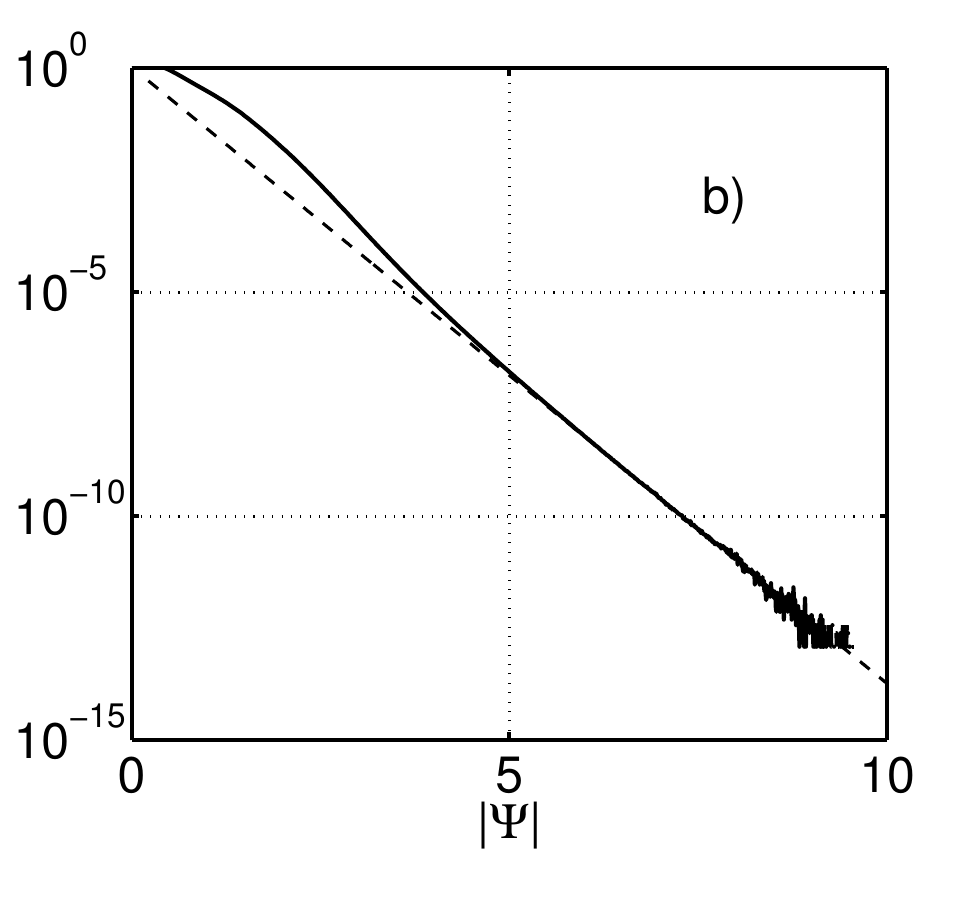}
\caption{\small {\it  (Color on-line) normalized squared amplitudes PDFs for generalized NLS equation accounting for six-wave interactions, dumping and pumping terms (\ref{Eq03}), $d_{l}=0.04$, $d_{2p}=0$, $d_{3p}=0.0004$, $p=0.05$, $\alpha=0.128$, in sem-logarithmic scale: versus $|\Psi|^{2}/\langle |\Psi|^{2}\rangle$ at $t=8$ (blue), $t=12$ (green), time-averaged PDF (red), time-averaged PDF of a linear system with the same spectra $I_{k}$ (black) as for Eq. (\ref{Eq03}) (a) and time-averaged PDF versus amplitude $|\Psi|$. Solid lines - PDFs, dashed lines - exponential dependence $\exp(-|\Psi|^{2}/2\sigma^{2})$, $\sigma\approx 0.75$, for graph (a) and $\exp(-|\Psi|/\Theta)$, $\Theta\approx 0.32$, tail for graph (b).}} 
\end{figure}

\begin{figure}[h] \centering
\includegraphics[width=130pt]{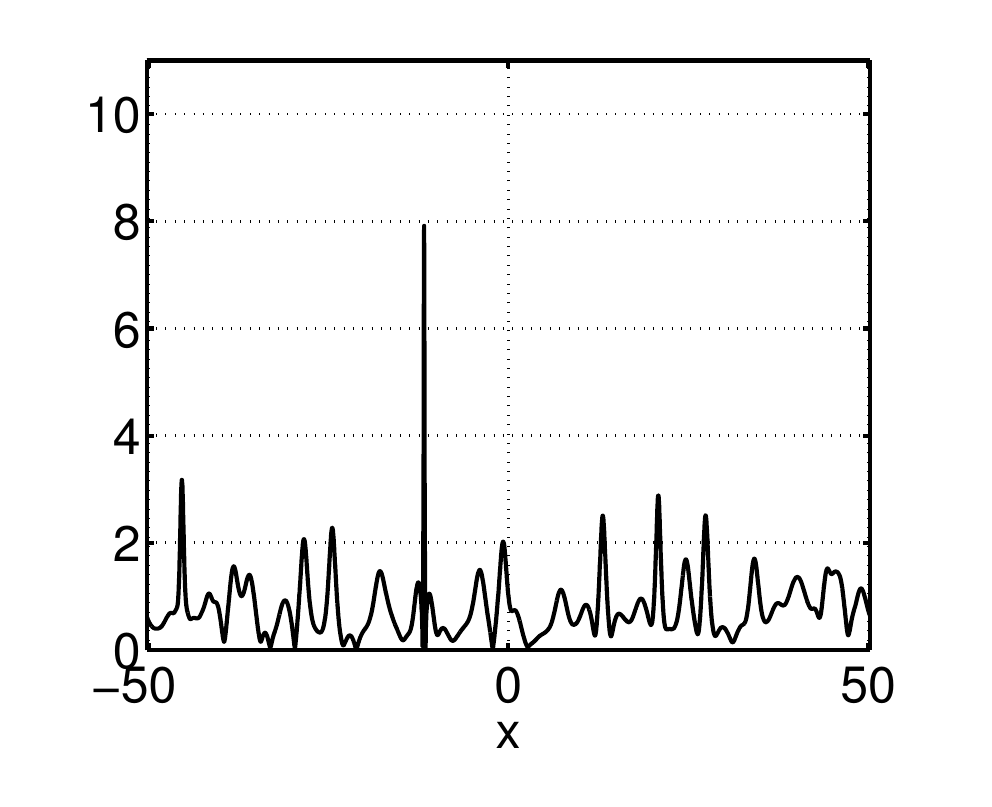}
\includegraphics[width=130pt]{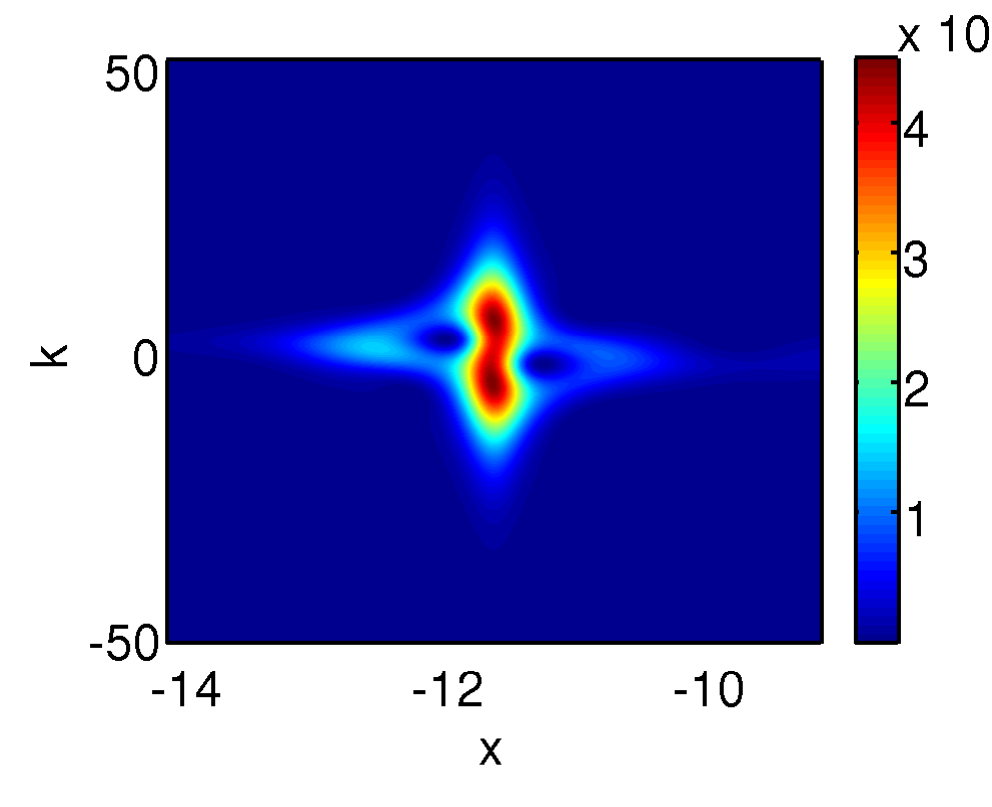}
\caption{\small {\it  (Color on-line) Field distribution $|\Psi|$ and spectrogram of a typical large wave event for generalized NLS equation accounting for six-wave interactions, dumping and pumping terms (\ref{Eq03}), $d_{l}=0.04$, $d_{2p}=0$, $d_{3p}=0.0004$, $p=0.05$, $\alpha=0.128$; $|\Psi|_{max}=7.9$ at $t=219.7$.}}
\end{figure}

\begin{figure}[h] \centering
\includegraphics[width=130pt]{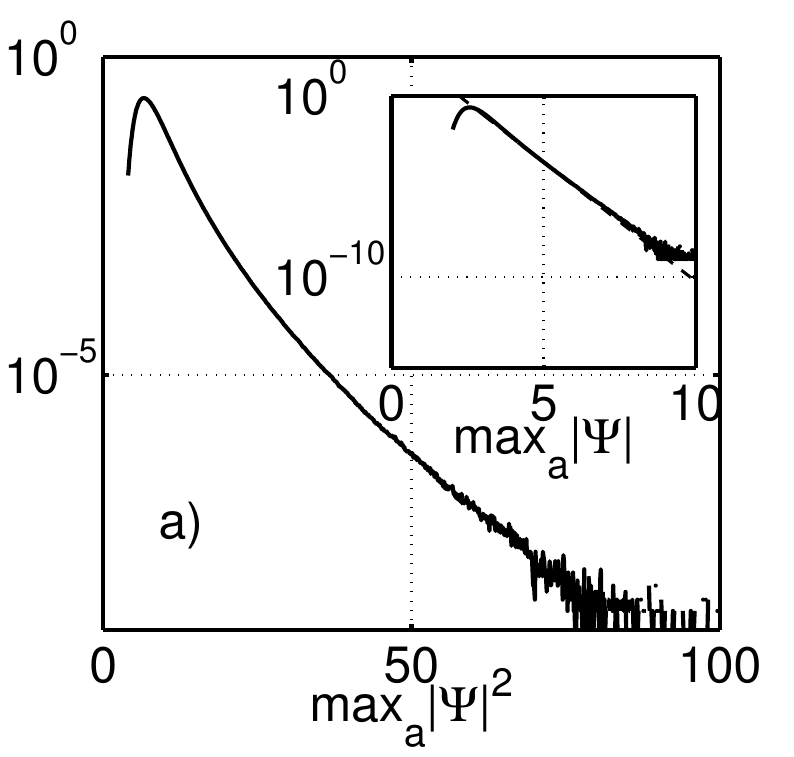}
\includegraphics[width=130pt]{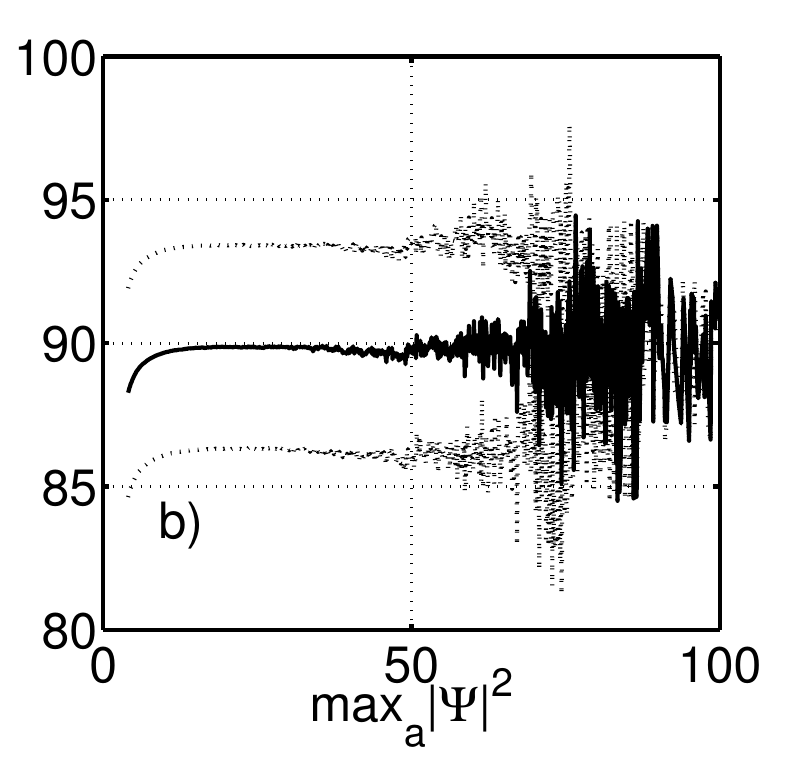}
\includegraphics[width=130pt]{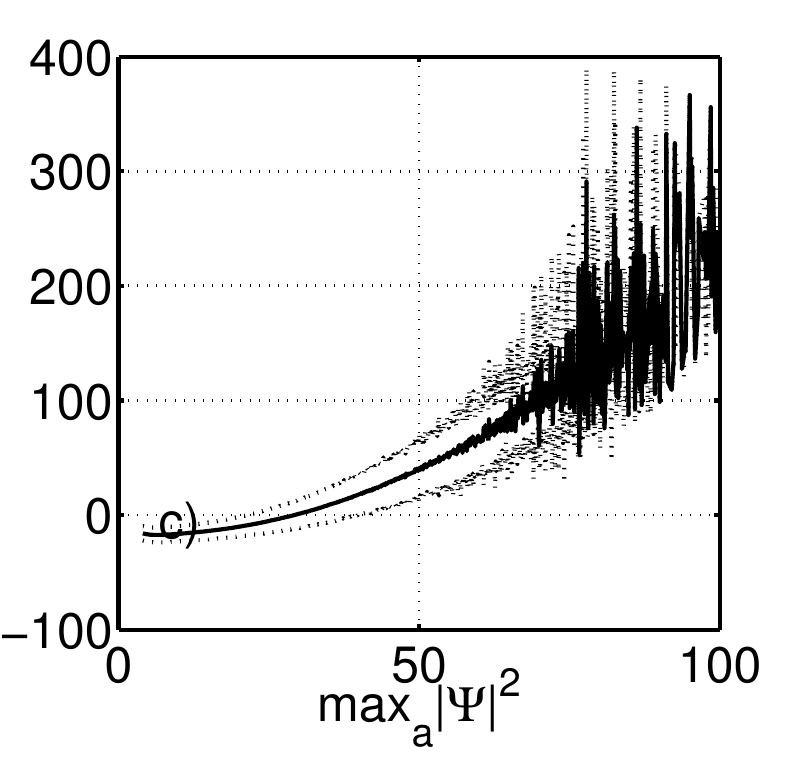}
\caption{\small {\it  Averaged over ensemble and time in the statistically steady state squared amplitudes and amplitudes (in-line) PDF for absolute maximums of wave field $\Psi$ (a), mean wave action $N$ (b) and mean total energy $H$ for distributions with the given absolute maximum $\max|\Psi|^{2}$ depending on $\max|\Psi|^{2}$ for generalized NLS equation accounting for six-wave interactions, dumping and pumping terms (\ref{Eq03}), $d_{l}=0.04$, $d_{2p}=0$, $d_{3p}=0.0004$, $p=0.05$, $\alpha=0.128$. Dashed line for in-line of graph (a) is non-Rayleigh tail $\exp(-|\Psi|/\Theta)$, $\Theta\approx 0.33$. Solid lines for graphs (b) and (c) are mean over ensemble values, dashed lines - borders for the corresponding standard deviations.}}
\end{figure}

We start numerical simulations from the same type of initial distributions $\Psi|_{t=0}=1+\epsilon(x)$, $|\epsilon(x)|\ll 1$ as in section 2a. Regularization of collapses performed in Eq. (\ref{Eq03}) with the proper choice of pumping and dumping coefficients allows one to constantly stay in the regime $|H_{6}|\ll |H_{d}|,|H_{4}|$, $|H_{d}|\sim|H_{4}|$, i.e. when dynamics of the system is close to that described by the classical NLS equation (see FIG. 14c). After the first 20-50 nonlinear lengths the system reaches statistically steady state when energy losses mainly as a result of regularization of collapses randomly appearing in space and time are compensated by the constant pumping of energy, and wave action, momentum and total energy as well as kinetic $H_{d}$, four- $H_{4}$ and six-wave interactions energy $H_{6}$ fluctuate near their mean values (see FIG. 14a,b,c). The statistically steady state is determined by coefficients $\alpha$, $d_{l}$, $d_{2p}$, $d_{3p}$ and $p$ only and does not depend on 
initial 
distribution $\Psi|_{t=0}$: thus, we observe the same results for initial distributions $\Psi|_{t=0}=\epsilon(x)$, $|\epsilon(x)|\ll 1$, with the exception of characteristic time necessary to reach the statistically steady state. In the steady state spectra $I_{k}$, spatial correlation functions $g(x)$, and the PDFs no longer depend on time that allows us to perform additional averaging over time.

In order to check the dependence of our results for averaged over ensemble spectra, spacial correlation functions and the PDFs on the specific values of pumping and dumping parameters we performed several numerical simulations for a set of different coefficients $d_{l}$, $d_{2p}$, $d_{3p}$ and $p$ with fixed six-wave coefficient $\alpha$ and found our results qualitatively do not depend on pumping and dumping parameters. Then we fixed $d_{l}=0.04$, $d_{2p}=0$, $d_{3p}=0.0004$ and $p=0.05$ and did five simulations for different six-wave interactions coefficient from $\alpha=0$ to $\alpha=0.256$. Corresponding spectra, spacial correlation functions and the PDFs in the statistically steady states are shown on FIG. 15. Thus, spectra consists of the upper exponentially decaying region $\,\sim \exp(-q_{1}|k|)$ corresponding to sufficiently small wavenumbers and the lower exponentially decaying region $\,\sim \exp(-q_{2}|k|)$, $q_{2}\le q_{1}$, corresponding to sufficiently large wavenumbers. Coefficient $q_{1}$ 
does not depend on six-wave coefficient $\alpha$, while $q_{2}=q_{1}$ when $\alpha=0$ and $q_{2}$ decreases with increasing $\alpha$. Spacial correlation functions approach to some universal form that for $x<x_{corr}$ is close to Gaussian and decays to zero as $x\to +\infty$. Typical evolution of the correlation length $x_{corr}$ is shown on FIG. 16.

The mean wave action at large time shifts changes with $\alpha$ from $N\sim 170$ at $\alpha=0$ to $N\sim 70$ at $\alpha=0.256$ because the main power drain from the system occurs during regularization of collapses \cite{Lushnikov}, while the number of collapses per time unit depends on $\alpha$: thus, six-wave interactions dominate for $\alpha=0.032$ starting from $|\Psi|>6$ and for $\alpha=0.256$ starting from $|\Psi|>2$. Since the mean wave action depends on six-wave interactions coefficient, mean squared amplitude $\langle |\Psi|^{2}\rangle$ also depend on it: $\langle |\Psi|^{2}\rangle$ decreases with increasing $\alpha$ from 1.7 at $\alpha=0$ to 0.7 at $\alpha=0.256$. 

Therefore on FIG. 15c we plot normalized squared amplitudes PDFs depending on $|\Psi|^{2}/\langle |\Psi|^{2}\rangle$ that allows us to examine PDFs for different $\alpha$, $d_{l}$, $d_{2p}$, $d_{3p}$ and $p$ all on one graph. The remarkable property of these PDFs is that 'fat' non-Rayleigh tails appear already in the region of medium amplitudes for non-zeroth six-wave interactions coefficient $\alpha>0$ and this non-Rayleigh addition increases with $\alpha$ that means the frequency of occurrence of large wave events increases with six-wave coefficient. On the other hand, in case of the absence of six-wave interactions $\alpha=0$ the corresponding PDF remains very close to Rayleigh distribution even for high waves, i.e. addition of pumping and dumping terms does not change the PDF (compare with section 2a). 

As shown on FIG. 17, in case of non-zeroth six-wave interactions $\alpha>0$ non-Rayleigh tails appear already in the nonlinear stage of modulation instability development and since then PDFs fluctuate near some universal form that turns out to be the same for all stages from modulation instability to the statistically steady state. The tails of the PDFs decay faster than any power of amplitude $|\Psi|$ and are similar to $\exp(-|\Psi|/\Theta)$ with some constant $\Theta$ as shown on FIG. 17b. Field distribution and spectrogram for a typical large wave event demonstrated on FIG. 18 as well as the corresponding motion dynamics show that extreme waves in case of Eq. (\ref{Eq03}) originate as a collision of several quasi-solitons that lead to appearance of collapses and their subsequent regularization.

Eq. (\ref{Eq03}) is a dissipative one with energy, wave action and momentum significantly varying with time. The latter means that different parts of the PDFs corresponding to small, medium and large amplitudes may be composed of distributions with significantly different integral characteristics. In order to check this we measured squared amplitudes PDF for absolute maximums $\max|\Psi|^{2}$ on one hand and mean wave action $N$ and total energy $H$ for the distributions with the given absolute maximum $\max|\Psi|^{2}$ depending on the value of absolute maximum $\max|\Psi|^{2}$ on the other hand. It turns our that PDFs for absolute maximums have the same properties as for field distributions $\Psi$: there is non-Rayleigh tail appearing already at medium amplitudes that decays as $\exp(-|\Psi|/\Theta)$, as demonstrated on FIG. 19a. Mean wave action $N$ for field distributions with the given absolute maximum does not depend on absolute maximum, therefore, all parts of the PDFs are composed of the distributions 
with the same wave action $N$ - or in other words with the same mean squared amplitude $\langle |\Psi|^{2}\rangle$ (FIG. 19b). It is very interesting that the mean total energy $H=H_{d}+H_{4}+H_{6}$ for the distributions with the given absolute maximum significantly increases with absolute maximum from negative values at small waves to large positive values at large waves (FIG. 19c), so that, taking into account inequalities $H_{4}<0$ and $H_{6}<0$, kinetic energy for distributions with extreme waves is higher than the potential one $H_{d}>|H_{4}+H_{6}|$. The latter one becomes possible at the expense of four-waves interactions energy $H_{4}$: on the extreme waves four-waves interactions become less important than the six-wave interactions, $|H_{4}|\ll |H_{6}|$, and six-wave interactions energy becomes comparable with kinetic energy $|H_{6}|\sim H_{d}$. Therefore, distributions with extreme waves that compose non-Rayleigh tails of the PDFs turn out to be significantly different from that with the ordinary 
waves in the sense of their integral characteristics: even though the mean squared amplitude $\langle |\Psi|^{2}\rangle$ is the same for both cases, total energy of such distributions is positive and significantly exceeds that for ordinary waves; and as might be expected, in case of extreme waves the dynamics of the system is close to that of the quintic NLS equation, $|H_{4}|\ll |H_{6}|$, $|H_{6}|\sim H_{d}$ even though the resulting PDFs are significantly different for these two equations (compare with \cite{Lushnikov}).\\


{\bf 4. Conclusions and acknowledgements.} \\

We would like to underline three our main results. First, we observe 'strange' results for the classical integrable NLS equation with condensate initial condition: there is a high peak at zeroth harmonic in averaged over ensemble spectra that fluctuates with time but never disappears, spacial correlation functions do not decay to zero level, there is a 'breathing' region on the PDFs for medium wave amplitudes and the frequencies of amplitude appearance significantly fluctuate with time. On the other hand for large amplitudes PDFs decay according to Rayleigh law. Addition of small dumping and pumping terms that breaks integrability leads to disappearance of the peak at zeroth harmonic in spectra, in this case spacial correlation functions decay to zero and in the statistically steady state PDFs turn out to be strictly Rayleigh ones.

It is necessary to note that even despite highly nonlinear regime we observed PDFs very close to Rayleigh ones for the classical NLS equation (\ref{Eq01}) except regions of quasi-periodicity points and also strictly Rayleigh PDFs for nonintegrable NLS equation accounting for small dumping and pumping terms (\ref{Eq01_1}). Therefore, presence of nonlinearity and significantly nonlinear regime of a system do not necessarily mean non-Rayleigh PDFs.

Second, we observed strongly non-Rayleigh PDFs for the classical NLS equation (\ref{Eq01}) with cnoidal wave initial condition. This result is very interesting in the sense that usually appearance of 'fat tails' on the PDFs is associated with nonintegrability of the system (see \cite{Hadzievski, Passot}). It turns out that the PDFs change to those close to Rayleigh ones as cnoidal wave's imaginary half-period $\omega_{1}$ increases from small values where cnoidal wave represents a lattice of nearly non-interacting thin and high solitons to high values where it approaches to condensate solution $F(x)=1/\sqrt{2}$. For initial conditions with significant overlapping of solitons inside the cnoidal wave we also observe peaks at zeroth harmonic in spectra, non-decaying to zero level spacial correlation functions and noticeably fluctuating with time frequencies of amplitudes appearance. 

Third, we demonstrated presence of non-Rayleigh tails $\,\sim|\Psi|\exp(-|\Psi|/\Theta)$ for amplitudes or $\,\sim\exp(-|\Psi|/\Theta)$ for squared amplitudes PDFs for generalized NLS equation accounting for six-wave interactions, pumping and dumping terms (\ref{Eq03}) even when six-wave interactions are small compared to four-wave interactions. The corresponding non-Rayleigh addition does not qualitatively depend on dumping and pumping parameters and also six-wave constant $\alpha$ until $\alpha>0$, increase with $\alpha$ and disappear in absence of six-wave interactions $\alpha=0$. It is interesting to note that for the condensate initial condition in general our results are similar to those for the statistics of rogue waves for Salerno model for discrete nonlinear lattices \cite{Hadzievski} and also for Alfven wave turbulence \cite{Passot} as we observe close to Rayleigh PDFs for integrable model Eq. (\ref{Eq01}) and appearance of strong non-Rayleigh tails when additional higher-order nonlinear terms are 
present in Eq. (\ref{Eq03}).

D. Agafontsev thanks E. Kuznetsov for valuable discussions concerning this publication, M. Fedoruk for access to and V. Kalyuzhny for assistance in author's work with Novosibirsk Supercomputer Center. This work was done in the framework of Russian Federation Government Grant (contract no. 0035 with Ministry of Education and Science of RF, November 25, 2010), and also supported by the program of Presidium of RAS "Fundamental problems of nonlinear dynamics", program of support for leading scientific schools of Russian Federation and RFBR grant 09-01-00631-a.\\


{\bf Appendix: numerical methods.} \\

We solve Eq. (\ref{Eq01})-(\ref{Eq03}) numerically in the box $-16\pi \le x<16\pi$ with periodical boundary conditions. In our numerical simulations we used the 2nd-order Split-Step method in which linear and nonlinear parts of the equations were calculated separately. In order to improve simulations and save computational resources we employed adaptive change of spacial grid size $\Delta x$ reducing it when Fourier components of solution $\Psi_{k}$ at high wave numbers $k$ exceeded $10^{-13}\max|\Psi_{k}|$ and increasing $\Delta x$ when this criterion allowed. In order to prevent appearance of numerical instabilities, time step $\Delta t$ also changed with $\Delta x$ as $\Delta t = h\Delta x^{2}$ with $h \le 0.1$ (see \cite{Lakoba}).

Implementation of such numerical schema allowed us to safely shift up to $t\sim 100$ nonlinear lengths, i.e. when 
\begin{equation}\label{criterion_single}
||\Psi_{h}(t,x) - \Psi_{0.5h}(t,x)||/||\Psi_{0.5h}(t,x)||\ll 1,
\end{equation}
where $\Psi_{h}(t,x)$ is the numerical solution calculated with a fixed coefficient $h=\Delta t/\Delta x^{2}$. We also did comparisons with 4th-order Split-Step method \cite{Muslu, Mclachlan} as well as with 4th- and 5th-order Runge-Kutta methods that confirmed validity of our results for \textit{single simulations} up to $t\sim 100$ nonlinear lengths.

Thus, all methods we used, including comparisons with the results obtained with higher number of Fourier modes (lower $\Delta x$) with the same initial noise, gave virtually the same results for the classical NLS equation with condensate initial condition up to time shifts $t\sim 60$. Beyond 60 nonlinear lengths all methods and all schema parameters $\Delta t$ and $\Delta x$ we used gave different results in the sense of criterion (\ref{criterion_single}). This behavior is connected with quasi-periodical dynamics of the classical integrable NLS equation: near $t\sim 60$ wave field again can be represented as $\Psi=1+\xi(x)$ with $|\xi(x)|\ll 1$, but this time $\xi(x)$ contains numerical errors that are unique for numerical method and it's parameters $\Delta t$ and $\Delta x$. Therefore, beyond 60 nonlinear lengths modulation instability develops differently for different numerical methods and their parameters.

Nevertheless, comparison of our \textit{statistical results} calculated with $h_{0}\sim 1/12$, in particular PDFs, revealed no difference with the results obtained with $h=0.25h_{0}$ or with the help of 4th-order Split-Step or 4th- and 5th-order Runge-Kutta methods far beyond 100 nonlinear lengths for all of the nonlinear systems considered in this publication.\\

\end{document}